\documentclass[sigconf]{acmart}

\usepackage[utf8]{inputenc}

\usepackage{soul}
\usepackage{url}
\usepackage{booktabs}
\usepackage{hyperref}
\usepackage{natbib}
\usepackage{amsmath,amsfonts}
\usepackage{algorithmic}
\usepackage{graphicx}
\usepackage{verbatimbox}
\usepackage{makecell}
\usepackage{textcomp}
\usepackage{xcolor}
\usepackage{hyperref}
\usepackage{xspace}
\usepackage{cleveref}
\usepackage{diagbox}
\usepackage[caption=false]{subfig}  
\usepackage{listings}
\usepackage{multirow}
\usepackage{placeins}
\usepackage{fancyvrb}
\usepackage{float}
\floatstyle{plaintop}
\restylefloat{table}

\newcommand{\ie}{\emph{i.e.,}\xspace}
\newcommand{\eg}{\emph{e.g.,}\xspace}

\newcounter{spcounter}

\begin{document}

\title{Out of the BLEU: How Should We Assess Quality of the Code Generation Models?} 

\author{Mikhail Evtikhiev}
\affiliation{
  \institution{JetBrains Research}
  \country{Republic of Cyprus}
 }
\email{mikhail.evtikhiev@jetbrains.com}

\author{Egor Bogomolov}
\affiliation{
  \institution{JetBrains Research}
    \country{Republic of Cyprus}
  }
  \email{egor.bogomolov@jetbrains.com}

\author{Yaroslav Sokolov}
\affiliation{
  \institution{JetBrains}
  \country{Germany}
  }
  \email{yaroslav.sokolov@jetbrains.com}

\author{Timofey Bryksin}
\affiliation{
  \institution{JetBrains Research}
    \country{Republic of Cyprus}
  }
  \email{timofey.bryksin@jetbrains.com}

\begin{abstract}
In recent years, researchers have created and introduced a significant number of various code generation models. 
As human evaluation of every new model version is unfeasible, the community adopted automatic evaluation metrics such as BLEU to approximate the results of human judgement.
These metrics originate from the machine translation domain and it is unclear whether they are applicable for the code generation tasks and how well they agree with the human evaluation on this task.
There are also other metrics, CodeBLEU and RUBY, developed to estimate the similarity of code, that take into account the properties of source code.
However, for these metrics there are hardly any studies on their agreement with the human evaluation.
Despite all that, minimal differences in the metric scores have been used in recent papers to claim superiority of some code generation models over the others.

In this paper, we present a study on the applicability of six metrics---BLEU, ROUGE-L, METEOR, ChrF, CodeBLEU, and RUBY---for evaluation of code generation models.
We conduct a study on two different code generation datasets and use human annotators to assess the quality of all models run on these datasets. 
The results indicate that for the CoNaLa dataset of Python one-liners, none of the metrics can correctly emulate human judgement on which model is better with $>95\%$ certainty if the difference in model scores is less than 5 points.
For the HearthStone dataset, which consists of classes of a particular structure, a difference in model scores of at least 2 points is enough to claim the superiority of one model over the other.
Our findings suggest that the ChrF metric is a better fit for the evaluation of code generation models than the commonly used BLEU and CodeBLEU. Yet, finding a metric for code generation that closely agrees with humans requires additional work.
\end{abstract}

\maketitle

 \section{Introduction}\label{sec:introduction}
Code generation systems are a way to make the process of writing source code easier and more accessible. 
In a common formulation, such systems take an intent---description in a natural language---as an input and produce a snippet of code that implements the intent.
Proper code generation is a long-standing problem~\cite{balzer1985} that, if implemented well, would aid in education, simplify drafting program implementations for non-programmers, and attract new programmers who may have limited programming experience in a given language~\cite{chen2021evaluating}. 
Therefore, having a strong code generation model could be very beneficial for the software development industry.

Currently, there are many various code generating models~\cite{Yin, GCNN, tranx, reranking, chen2021evaluating} and several datasets~\cite{spider, oda, agashe, card2code, Docstrings, conala, lu2021codexglue} on which these models are evaluated. 
The code generation models are usually assessed with either accuracy, BLEU metric~\cite{BLEU}, or CodeBLEU metric~\cite{CodeBLEU}. 
Originally, BLEU was created to evaluate the quality of machine translation for natural language processing, and it was empirically validated to be correlated with the human judgments of the translation quality for natural language texts. 
However, no such validation exists for the code generation task. 
Moreover, for the closely related code migration problem, Tran et al.~\cite{RUBY} have shown that the BLEU results are only weakly correlated with the human judgment. 
For the related code summarization problem, Roy et al.~\cite{roy2021reassessing} have shown that BLEU metric is a less reliable indicator of human judgement than other metrics, such as METEOR or ChrF.

We identify three possible problems with the application of the BLEU metric for the code generation task, that, to the best of our knowledge, have hardly been addressed~\cite{RUBY, CodeBLEU}:
\begin{itemize}
    \item  It is unclear whether existing metrics are suitable for the assessment of the code generation models. 
    \item It is unclear how significant the metrics scores are and how big the difference in the scores should be to claim one model's superiority over the other. 
    \item It is unclear how well the metrics correlate with the human judgement for existing code generation datasets. 
\end{itemize}

In our study, we consider two different datasets. 
The CoNaLa dataset~\cite{conala} is a dataset of questions posted on Stack Overflow\footnote{Stack Overflow:~\url{https://stackoverflow.com/}} with solutions in Python. 
The solutions are short and generally are one line long.
Card2code Hearthstone~\cite{card2code} is a dataset dedicated to generating classes that are descriptions of the cards used in the Hearthstone game.
The classes are rigid and most of the class structures are identical for every snippet.
For each of the datasets, we consider several machine learning models for code generation. 

For the CoNaLa dataset, we compare the results of five different models: 1) CoNaLa baseline~\cite{conala}, 2) Codex~\cite{chen2021evaluating}, 3) TranX without pretraining~\cite{tranx}, 4) TranX with pretraining, and 5) TranX with pretraining and reranking~\cite{reranking}. 
While being publicly available, the selected models greatly vary in quality and complexity, which allows for judgement on the relation between the models' quality, metric values, and human assessment. 
For the Hearthstone dataset, we compare the results of two models that were previously evaluated on this dataset: NL2Code~\cite{Yin} and GCNN~\cite{GCNN}. 

To address the problem of the applicability of automated metrics, we carry out paired bootstrap resampling~\cite{efron1983estimating}.
We consider BLEU, METEOR, ROUGE-L, ChrF, CodeBLEU, and RUBY~\cite{BLEU, ROUGE, METEOR, RUBY, CodeBLEU, popovic2015chrf} metric scores of the models. 

To address the problem of correlation between human assessment and computer metric scores, we carry out a human evaluation of the generated snippets. 
Software developers evaluated whether the suggested snippets were helpful in solving the posed problem on the scale from 0 to 4. 
For the CoNaLa dataset, 12 developers took part in the evaluation and we got on average $4.5$ grades from different developers per snippet. 
For the Hearthstone dataset, there were four graders, and every grader evaluated the entire dataset.

The amount of grades per snippet we collect is not enough to analyze the metrics performance on the snippet level, as Mathur et al.~\cite{mathur2020tangled} argued it is necessary to have 15 grades per snippet to provide a stable score. 
Thus, we focus on the comparison of models at the corpus level. 
The available set of ML models is not large enough to study the significance of difference in metric scores: for example, for the CoNaLa dataset there only are five original models, and thus only ten different pairs of models to compare. 
To provide a statistical analysis of corpus-level score differences, we augment the original set of models with a set of synthetic models. 
In it, we replace a part of some model predictions with predictions that have a higher or lower human assessment score, following Roy et al.~\cite{roy2021reassessing}.

Our findings and contributions are the following:
\begin{itemize}
    \item We find that the existing metrics are not suitable for assessing code generation, as for every dataset and every metric, metrics disagree with the human judgement in more than 5\% of the cases.
    \item We find that the difference in metric scores of two models of less than two points on a 0--100 scale is statistically insignificant in more than 5\% of the cases. 
    This finding does not depend on the human evaluation and shows it is necessary to test for statistical significance when reporting increase in metric scores of less than two points.
    \item We find that, when taking human assessment into account, all metrics are unreliable on the CoNaLa dataset if the score difference is less than five points, and are unreliable on the HearthStone dataset if the score difference is less than two points. 
    Of all metrics we considered, ChrF and ROUGE-L are the best-performing metrics for the code generation task. 
  
\end{itemize}

This paper is structured as follows. 
In Section~\ref{sec:background}, we describe the code generation problem, briefly describe the metrics we use for assessing generated code, and describe a similar study by Roy et al.~\cite{roy2021reassessing} that targeted the code summarization problem.
In Section~\ref{sec:Motivation}, we compare usage of automated metrics with test-based evaluation, and outline possible issues with the current usage of automated metrics.
In Section~\ref{sec:approach}, we describe the methodology of our study: outline the study pipeline, explain our choice of datasets and models, research questions, our approach to answering them.
In Section~\ref{sec:results-discussion}, we present our results and answer the RQs presented in the previous section.
In Section~\ref{sec:implications}, we summarize our findings to provide the guidelines for the practitioners who want to use autmoated metrics to assess code generation models, and present the directions for the future work.
In Section~\ref{sec:threats-to-validity}, we address the threats to the validity of our study.
In Section~\ref{sec:conclusions}, we summarize our paper.
In Appendix~\ref{appendix} we describe the metrics we study in more details.
Finally, our replication package can be found at \url{https://github.com/JetBrains-Research/codegen-metrics}.

\section{Background}\label{sec:background}

\subsection{Code Generation}
Code generation is a long-standing problem~\cite{balzer1985}, and a good code generation model could decrease the barrier for writing code, automate some of the routine tasks engineers have, and help non-programmers create programming solutions for their problems.
This problem is also related to other applications of machine learning to code.
In the greater context of code-related tasks, code generation is a task complementary to code summarization and is closely related to code migration and code completion. 

The development of deep learning has enabled the successful application of various neural models to the code generation problem.
In particular, Ling et al.~\cite{card2code} suggested a sequence-to-sequence model to generate code from natural language descriptions. 
Yin et al.~\cite{Yin} and Rabinovich et al.~\cite{Rabinovich} modified the standard decoder that generates a sequence of tokens to enforce grammar rules by first generating an abstract syntax tree and then converting it into code. 
Sun et al.~\cite{GCNN} suggested replacing recurrent neural networks with grammar-based structural convolutional neural networks. 
Unlike recurrent neural networks, convolutional neural networks can track the context even between distant regions of the analyzed data. 
In contrast, recurrent neural networks are not capable of tracking the context when relevant pieces of information are far apart, also known as the long dependency problem~\cite{hochreiter, bengio}. 
Wei et al.~\cite{Wei} suggested dual training of code generation and code summarization models to enhance the quality of both models.

In contrast to recurrent neural networks, models based on the Transformer architecture~\cite{vaswani2017} process the whole sequence simultaneously, which is more efficient both in terms of computational speed and capturing the dependencies between distant tokens. 
Nowadays, we observe rapid progress in the quality of code generation models due to gigantic Transformer-based models such as Codex~\cite{chen2021evaluating}, AlphaCode~\cite{alphacode}, and CodeParrot.\footnote{CodeParrot's page on HuggingFace: \url{https://huggingface.co/codeparrot/codeparrot}} 

Table~\ref{tab:cg-papers} summarizes types of neural networks and metrics used by the researchers in the papers discussed above.

\begin{table*}[]
\begin{tabular}{llll}
\toprule
\multicolumn{1}{c}{\textbf{Paper}}             & \multicolumn{1}{c}{\textbf{NN type}}            & \multicolumn{1}{c}{\textbf{Metrics}}                                  & \multicolumn{1}{c}{\textbf{Year}} \\ \midrule
Barone et al.~\cite{Docstrings}     & NMT                & BLEU                                     & 2017 \\
Chen et al.~\cite{chen2021evaluating}        & Transformer        & BLEU, Pass@k                             & 2021 \\
CodeParrot        & Transformer        & Pass@k                             & 2021 \\
AlphaCode~\cite{alphacode}        & Transformer        & Evaluation on Codeforces                             & 2022 \\
Ling et al.~\cite{card2code}        & RNN                & BLEU, Accuracy                           & 2016 \\
Lu et al.~\cite{lu2021codexglue}         & RNN, Transformer   & BLEU, Accuracy, CodeBLEU                 & 2021 \\
Rabinovich et al.~\cite{Rabinovich} & RNN                & BLEU, Accuracy, F1                       & 2017 \\
Ren et al.~\cite{CodeBLEU}        & PBSMT, Transformer & BLEU, Accuracy, CodeBLEU                 & 2020 \\
Sun et al.~\cite{GCNN}        & CNN, RNN           & Accuracy, BLEU                           & 2019 \\
Wei et al.~\cite{Wei}        & RNN                & BLEU, Percentage of valid code           & 2019 \\
Yin et al.~\cite{Yin}        & RNN                & BLEU, Accuracy                           & 2017 \\
Yin et al.~\cite{tranx}        & RNN                & \makecell[l]{Execution accuracy, \\exact match accuracy} & 2018 \\
Yin et al.~\cite{reranking}        & RNN                & BLEU, Accuracy                           & 2019\\\bottomrule
\end{tabular}
\caption{\label{tab:cg-papers}Comparison of various code generation papers}
\end{table*}

\subsection{Evaluation of Code Generation Models}

To be able to track improvements of a model, it is necessary to evaluate its performance.
Human assessment is the gold standard for most machine translation or machine generation problems. 
However, manual assessment is also very expensive and slow, and it is impractical to do human evaluation for each generated sample during the model development. 
Thus, it is crucial to have an easy to compute metric to evaluate the output of a model.

The code generation task is no different. The evaluation approaches for code generation can be split into three categories:
\begin{enumerate}
    \item Metrics from the machine translation domain;
    \item Metrics developed to compare code snippets;
    \item Running and testing the generated code.
\end{enumerate}

Further, we discuss all three in detail.

\subsubsection{Metrics from machine translation}
As \Cref{tab:cg-papers} shows, the quality of code generation models is typically assessed by the BLEU metric score~\cite{BLEU} or accuracy. 
The BLEU (BiLingual Evaluation Understudy) metric is a metric that was originally developed for the automatic quality evaluation of machine-translated texts.
The BLEU metric is a corpus-level metric based on the modified $n$-gram precision measure with a length penalization for the candidate sentences that are shorter than the reference ones.

Researchers also consider other machine translation metrics: 
\begin{itemize}
    \item ROUGE-L~\cite{ROUGE} is a recall-oriented metric that looks for the longest common subsequence between the reference and the candidate. 
    \item METEOR~\cite{denkowski2014meteor} is a mixed recall-precision metric that also penalizes candidates for not having adjacent unigrams that are adjacent in the reference example. 
    \item ChrF~\cite{popovic2015chrf} is a character n-gram F-score metric, where precision and recall in the F-score computation are averaged over 1- to 6-grams of characters.
\end{itemize}

In addition to the aforementioned metrics, researchers often report Accuracy as an additional metric. 
While it supports the fact that one model is superior to another, it is rarely used as the primary metric in the generation tasks due to being too strict and less robust. 
Thus, we do not analyze Accuracy in our study, as we focus on metrics used for direct model comparison.

\subsubsection{Metrics designed for code}
~\par \textbf{RUBY.} The RUBY metric was suggested by Tran et al.~\cite{RUBY} as an alternative to the natural languages metrics.
Indeed, even though BLEU (like METEOR and ROUGE-L) was originally created for the assessment of machine translation models for natural languages, it is widely used for assessing code generation, code migration, and code summarization models. 
Tran et al.~\cite{RUBY} conducted an empirical study on BLEU to check its suitability in the context of the code migration task. 
In their paper, they show that the BLEU metric has a rather weak correlation of 0.583 with the human assessment. 
The authors also constructed a synthetic dataset to illustrate that BLEU may yield similar results for the models whose quality differs from the perspective of the human grader. 
To address this issue, the authors devised a new metric RUBY, which takes code structure into account. 
The metric compares program dependency graphs (PDG) of the reference and the candidate; if a PDG is impossible to build, it falls back to comparing abstract syntax trees (AST), and if an AST is also impossible to build, the metric compares the weighted string edit distance between the (tokenized) reference $R$ and candidate sequences. 

\textbf{CodeBLEU.} Ren et al.~\cite{CodeBLEU} suggested a new metric called CodeBLEU to evaluate the quality of generated code for code generation, code translation, and code refinement tasks.
CodeBLEU is a composite metric with the scores being weighted average of 4 different sub-metrics treating code differently: as a data-flow graph, as an abstract syntax tree, and as text. 
For text, CodeBLEU provides two different submetrics, one of which treats all tokens as being equally important, and another gives higher weight to the keywords.

\subsubsection{Test-based evaluation}

The impressive performance of recent large-scale models~\cite{chen2021evaluating,alphacode} allows the use of evaluation techniques which are closer to practical applications: actually running the generated code on pre-written unit-tests and checking whether it solves the posed problem. 
For example, the authors of Codex~\cite{chen2021evaluating} also present a dataset called HumanEval which consists of programming tasks and tests validating the correctness of the generated code. 

While this approach is reasonable, we argue that for now it will not fully replace existing evaluation techniques that rely on the usage of automated metrics. 
In order to apply test-based evaluation, researchers need carefully created datasets for each particular code generation setting. 
Additionally, the studied models should pass large enough number of tests in order to robustly distinguish between them.

\subsection{Study of Metrics for Code Summarization}\label{sec:roy}

The automated metrics are used for a variety of other code-related generation tasks such as code translation, code summarization, or code refinement~\cite{lu2021codexglue}. 
Recently, Roy et al.~\cite{roy2021reassessing} studied the applicability of automated metrics for the code summarization task, which is closely related to code generation.
For this task, metrics such as BLEU are used widely as proxies of human evaluation.
The authors show that there is no statistically significant difference between the models with corpus scores different by less than 1.5 points according to any of the considered metrics.
Moreover, all the metrics the authors considered are not reliable proxies of human evaluation if the difference in corpus scores is less than two points according to the metrics.
Of all the metrics considered by Roy et al., METEOR, ChrF, and BERTScore show the best agreement with the human judgement on the corpus level.
As Roy et al. do an extensive study of the metric performance for a task that is closely related to the code generation, we adopt many of the methods they employed in our research.

\subsubsection{Dataset and labeling}
Roy et al. use the Java code summarization dataset of LeClair et al.~\cite{leclair2019recommendations}. 
They randomly sample 383 snippets from it and generate five summaries with different models.
Human annotators then evaluate the five generated summaries and the reference summary on a five-point Likert scale to assess the conciseness, fluency, and content adequacy of each summary. 
They also assign a Direct Assessment (DA) score on a 0--100 scale that reflects their opinion about the general quality of a summary.
Only the Direct Assessment score is used to analyze the relative metric performance.

\subsubsection{Corpus-level metric assessment}
The corpus-level assessment of metrics applicability by Roy et al. pursues two slightly different goals.
First, the authors are interested in whether the metrics are capable of distinguishing the quality of the existing models. 
To do that, they carry out randomized significance testing on the 383-snippet dataset to find that out of five models considered in the study, the difference in scores of the best five models is not statistically significant. 
It is important to highlight that this lack of statistical difference was found solely from the metric scores and does not rely on human labelling.

The second goal for the corpus-level metric assessment is to find whether the commonly used corpus-level metrics reflect human
quality assessments of generated summaries.
There is a relative shortage of available machine learning models (Roy et al. used five code summarization models in their study).
Thus, it is impossible to study directly what difference in metric scores is necessary to claim that one model is better than the other according to humans -- there are not enough pairs of models to get enough data on the differences in model scores.
However, if there were many more independent models, the researchers would have to label much more model outputs, increasing the cost and laboriousness of the study. 
In order to get more diversity in metric scores without increasing the number of summaries to label, Roy et al. use synthetic models. 

A synthetic model is a model that yields a set of summaries based on one of the five original models, with a varying proportion of summaries replaced by the predictions of the other models. 
In particular, to create a synthetic model that improves the original model A by 1\%, the authors replace 1\% of the summaries predicted by the model by the better predictions of other models.
The quality of the prediction is assessed according to the human DA score.
Roy et al. create a set of synthetic models and then select 100 of them.
Then, they add them to the five original models and do a pairwise comparison into several different buckets based on the statistical significance of the metric score difference as well as of the difference magnitude.
The bucket can be defined, for example, for statistically significant metric differences between two and five. 
For each of these pairs, Roy et al. also calculate the significance of the difference in their corresponding human DA scores. 
The effectiveness of a corpus-level metric can then be determined by looking at the agreement between the metric score and human assessment score. 
For a reliable automatic evaluation metric, one expects to find a one-to-one correspondence between significant differences in metric scores and human assessment scores.

Using the pairwise comparison approach, Roy et al. are able to analyze the following:
\begin{itemize}
    \item They find for how many pairs in a given bucket the two models in the pair are significantly different, according to each metric. 
    This allows to deduct what difference in the metric scores of two models outputs is necessary to expect so that the two models will also be significantly different from the metric's point of view.
    \item For each bucket and for every metric, they consider the group of pairs, in which one model is significantly better than the other according to the metric. 
    Then, for each pair, they check whether the two models in it are also significantly different, according to the human assessment.
    This allows them to study the Type-I error of each metric and check how it changes from bucket to bucket.
    \item For each bucket and for every metric, they consider the group of pairs, in which the two models are not significantly different, according to the metric. 
    Then, for each pair from this group, they check whether the two models in it are significantly different, according to the human assessment.
    This allows them to study the Type-II error of each metric and check how it changes from bucket to bucket.
\end{itemize}

From this analysis, Roy et al. find that automatic evaluation metrics are not able to accurately capture the differences in summarization quality between two approaches when the metric difference is less than two points.
METEOR, BERTScore, and ChrF perform the best in terms of Type-I and Type-II error rate. 
BLEU has the highest Type-I error rate regardless of the magnitude of the difference.

\subsubsection{Snippet-level analysis}
Roy et al. also consider the metric performance for the snippet level.
In principle, snippet-level metric result analysis can provide an advantage over corpus-level analysis by tracking fine-grained performance of the models.
To carry out the snippet-level analysis, Roy et al. use the Direct Assesment Relative Ranking technique, which compares the pairwise relative scores of two snippets~\cite{ma2019results}. 
This technique relies on the Direct Assessment scoring and cannot be applied to the annotations on the five-point scale.

\section{Motivation}\label{sec:Motivation}

Metrics are used during the validation phase of a machine learning pipeline and to compare different models. 
However, if human assessment is the golden standard, the used metric should align with the human judgement as closely as possible. 
For example, in machine translation, there is an annual contest between various metrics, with the best metric being the one that emulates human judgement the best~\cite{ma2018results, ma2019results}.

Even if some metric (such as BLEU) has been used in the past to emulate human judgement, it may be beneficial to consider other metrics which may have better correlation with human assessments. 
A similar situation has emerged in the natural language generation: even though BLEU was initially adopted to this domain, it was later shown that word-overlap based metrics (such as BLEU) have very low correlation with human judgement in certain natural language generation tasks such as dialog response generation~\cite{liu}.

In the rest of this section, we discuss in detail why studying the automated metrics for code generation is important and which questions are worth being answered in this regard.

\subsection{Metrics and Test-based Evaluation}

With the recent introduction of HumanEval~\cite{chen2021evaluating}, a dataset that allows running and testing generated Python code in a close-to-practical setting, it might seem that the usage of automated metrics will soon become obsolete. 
However, we think that it will not be the case in the near future.

Firstly, collecting test-based evaluation datasets requires significant human effort to develop a set of tasks as well as cover them with tests. 
Given that the code generation task can be formulated differently and applied to different languages and domains, each particular case requires a separate manually crafted evaluation system. 
Thus, the usage of automated metrics is helpful when adopting code generation in new domains.

Secondly, training and inferencing very large models such as Codex is both costly and technically challenging~\cite{chen2021evaluating}.
For this reason, an important direction of research is the development of smaller code generation models which cannot yet achieve the quality comparable to large Transformer-based counterparts. 
For smaller models, evaluation frameworks like HumanEval would lead to poor metric scores, and their robustness for model comparison in this case remains an open question. 

Finally, even if two models generate code that does not pass any tests, it still might be possible to say which piece of code is closer to the correct solution.
For example, for a problem ``Get rid of None values in dictionary \textit{d}'' and two pieces of code presented below the first piece is much closer to the right solution, even though it still does not pass the tests. 
\begin{verbatim}
    1. print(dict((k,v) for k,v in d.items() if v)))
    2. list(d.values())
\end{verbatim}
Thus, it is important to be able to evaluate the quality of generated code snippets even if they do not pass the tests, as developers might find some generated snippets easier to fix and integrate into their code.

\subsection{Are Existing Metrics Suitable for Code Generation?}
Machine translation metrics were developed for natural languages and do not take into account the properties of programming languages. 
The usage of such metrics might be sub-optimal for the code generation assessment due to several factors.

\subsubsection{Differences between programming and natural languages}
Programming languages have a strict syntactic structure, while the natural language structure is more relaxed. 
For example, while swapping two groups of tokens in a natural language sentence often does not strongly affect its meaning, such a transformation will often make a code snippet invalid. 
Secondly, machine translation (MT) metrics measure the lexical precision of the model output, while for the generated code we want to assess its functionality.

It is possible to make MT metrics somewhat more code-friendly, \textit{e.g.}, it is possible to rename all the variables in the candidate and the references according to their order of appearance, removing the spurious mismatch due to the different naming conventions. 
Yet, some issues cannot be apparently addressed without taking code structure into account. 
It is therefore plausible that a metric that will take into account the code snippets' structure and syntax will be a better proxy of the human assessment.

\subsubsection{BLEU has been outperformed in other tasks}
Human assessment is the best option for evaluating quality of a code generation model and is considered to be ground truth in metrics evaluation in many different tasks, see e.g.~\cite{reiter2018structured}.
However, as human evaluation is very expensive, it is obviously impossible to have every new output of the model evaluated by a group of programmers. 
A priori, it is unclear whether BLEU or any other metric scores are correlated well with the human assessment for the code generation task. 
Original papers for machine translation metrics~\cite{BLEU, ROUGE, METEOR, popovic2015chrf} include studies that show a high correlation between the metrics scores and the human judgement for the machine translation task.
However, a review by Reiter~\cite{reiter2018structured} shows that the BLEU–human correlations are poor for natural language generation tasks and BLEU should only be used to evaluate machine translation NLP systems.

For the closely related problem of code migration, it was shown that the correlation between BLEU scores and human grades is 0.583, which is rather weak~\cite{RUBY}.
There is also a study on the metric-human correlation for BLEU, accuracy, and CodeBLEU metrics~\cite{CodeBLEU}, which has shown that the CodeBLEU metric is better correlated with human opinion than accuracy or BLEU. 
However, this study did not consider other metrics.
Finally, Roy et al.~\cite{roy2021reassessing} did an extensive study on the applicability of the automated metrics for the code summarization problem, to find that the de-facto standard BLEU metric is one of the worst metrics for assessing code summarization models out of the six metrics they considered. 

All these observations highlight that the applicability of a particular metric strongly depends on the problem. 
Thus, using a metric that successfully works for one problem for another problem may be unwarranted.

\subsubsection{Translation from metrics to human assessment}
It is unclear that an increase in a metric score is linearly related to the increase of the ``true'' quality of the code snippet. 
For an illustration, let us consider one of the tasks in the CoNaLa dataset:
\begin{Verbatim}[commandchars=\\\{\}]
\textbf{Task:} concatenate a list of strings ['a', 'b', 'c']
\textbf{baseline model solution:} set(['a','b','b'])
\textbf{best-tranx-rerank solution:} ''''''.join(['a','b','c'])
\end{Verbatim}

Even though the baseline snippet fails to solve the task question (and didn't even manage to reproduce the list of strings that need to be concatenated), it has a relatively high BLEU score of 48.09. 
The second snippet successfully solves the problem and has a BLEU score of 100. 

Now, let us consider hypothetical outputs of two different models A and B. 
Both outputs have BLEU 50, but for model A every candidate has BLEU 50 and is of quality similar to the one above, while for model B, half of the candidates have BLEU zero and the other half have BLEU 100.
In this case, it may be argued that model B is better than model A, even if they have close corpus-level BLEU scores: given the example above, model A can generate hardly relevant code snippets all the time, while model B generates perfect code in half of the cases.

If the dependency between human assessment and metric values is not linear, we cannot simply average the metric values over all the snippets to reflect the human assessment of the model. 
In addition, there might be other reasons why BLEU scores and human scores might not correlate well, and it is necessary to study the correlation between the two to be able to infer the knowledge how to interpret BLEU scores and assess the models' quality from them. 

\subsection{Do We Use Automated Metrics Correctly?}

The common way of using automated metrics to assess models is to report a single corpus-level number~\cite{Yin, Rabinovich, tranx, lu2021codexglue}. 
While this approach is simple and might be very practical during the training process, it is unclear how the raw difference in metric scores can be translated into statements on the statistical significance of the difference. 

In the code generation domain, comparison of different models is usually done by simply comparing their BLEU or CodeBLEU scores, averaged over the entire test dataset (see \eg~\cite{Yin, Rabinovich, tranx, lu2021codexglue}). 
However, when an improvement from \eg the BLEU score of 29 to the BLEU score of 30 is claimed, it is rarely supported by data on the statistical significance of the improvement.
As Roy et al.~\cite{roy2021reassessing} have shown, for the closely related code summarization task, small differences in metric scores are statistically insignificant, it is possible that the same phenomenon exists for code generation.

Therefore, it is important to study how big the difference between the metric scores of two models for a particular dataset should be to claim that one of the models is better than the other with the desired confidence.

\section{Methodology}\label{sec:approach}
The problems we list in \Cref{sec:Motivation} have motivated us to pose the following research questions:
\begin{itemize}
    \item[RQ1] Does the performance of the considered models differ significantly on the corpus level?
    \item[RQ2] How significant are the results of automated metrics and how big should be the difference in corpus-level metric scores of two models to claim that one model is better (according to the given metric) than the other with predefined significance? 
    \item[RQ3] How well do the corpus-level metric scores reflect the human assessment of generated code?
\end{itemize}

\begin{figure*}[t]
\includegraphics[width=\textwidth]{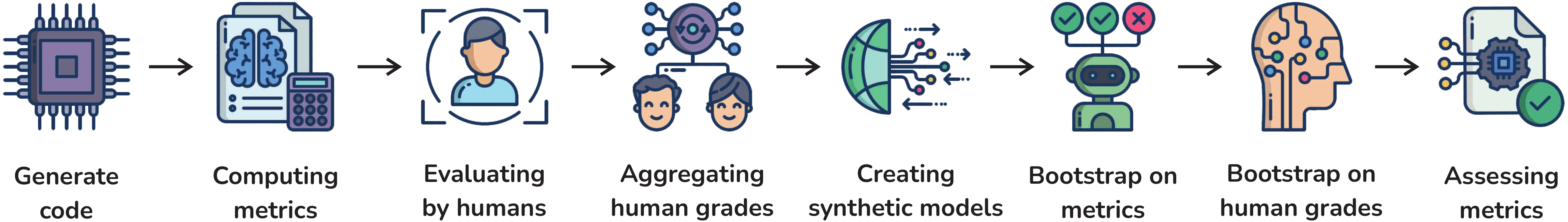} \caption{A high-level description of the pipeline of our approach.} \label{fig:pipeline}
\end{figure*}

Inspired by the work of Roy et al.~\cite{roy2021reassessing} described in detail in \Cref{sec:roy}, the pipeline of our approach is as follows:
\begin{itemize}
    \item[1.] We collect the models' output on the datasets we consider.
    \item[2.] We evaluate automated metrics on the generated code snippets, getting every metric score for every generated snippet.
    \item[3.] We carry out a human evaluation of the generated snippets (described below in more details), collecting a set of human grades for every generated snippet.
    \item[4.] Using the obtained set of human grades, we get the ``ground truth'' human grade by aggregating the grades together with the M-MSR algorithm~\cite{ma2020adversarial}, getting a single grade for each snippet evaluated by the experts. We use the realization of the M-MSR algorithm by Ustalov et al.~\cite{HCOMP2021/CrowdKit}.
    \item[5.] Using the models' output, we create synthetic models by replacing some of the predictions with the predictions that received higher or lower human assessment score. 
    For example, to get a synthetic \texttt{tranx-annot} model with 1\% of predictions improved, we consider its outputs and replace 1\% of its worst predictions with the best predictions available from other models.
    The quality of a prediction is derived from the human assessment score.
    \item[6.] For every pair of both synthetic and non-synthetic models evaluated on the same dataset, we carry out paired bootstrap resampling. 
    We do that to find the statistical significance of the claim that one of the models is better than the other according to the metric scores. 
    We use a $95\%$ threshold to claim a statistically significant difference between the models.
    \item[7.] For each dataset evaluated by humans and for every pair of models evaluated on it, we carry out paired bootstrap resampling on the ground truth grades. 
    We use the statistical test results to check with what statistical significance we can infer that one of the models is better than the other according to the human opinion.
    \item[8.] Following Mathur et al.~\cite{mathur2020tangled} and Roy et al.~\cite{roy2021reassessing}, we carry out a pairwise model comparison of human assessment and corpus level metrics for CoNaLa and Hearthstone datasets. 
    We start by computing the difference in corpus-level metric scores for all pairs of models evaluated over the given dataset. 
    We then divide these model pairs into several buckets according to the difference in the metric scores; we also have an extra bin for the pairs which metrics cannot distinguish.
    For each of the pairs in every bin, we check if the human evaluation agrees with the metric evaluation, \ie whether humans distinguish the pair of models or not.
\end{itemize}
It would be interesting to carry out a comparative analysis of metrics on the snippet level.
However, Mathur et al.~\cite{mathur2020tangled} argued that it is necessary to collect at least 15 human assessments per snippet in order to provide a stable score and analyze metrics performance on the snippet level. 
As we were able to collect four grades per snippet for the Hearthstone dataset and 4.5 grades per snippet for the CoNaLa dataset, we opt not to analyze metric performance on the snippet level.

\subsection{Datasets and Models}
In our study, we consider two different datasets: CoNaLa~\cite{conala} and Card2code Hearthstone~\cite{card2code}. 
We focus on the datasets containing general Python code, leaving the non-Python datasets such as Spider (containing SQL)~\cite{spider} and JuICe (containing Jupyter Notebooks)~\cite{juice} out of the scope.
We also leave out the CodeXGLUE dataset~\cite{lu2021codexglue}, since the code to text problems in CodeXGLUE dataset come from the Concode dataset~\cite{iyer2018mapping}, which is focused on the Java code. 
Given the high popularity the CodeXGLUE dataset has gained as of late, it would be interesting to extend our research later to this dataset.

For both datasets, we use the models suggested by authors in previous works. We employ the original implementations, hyperparameters, and---if possible---the original trained weights or the code generated by the model as provided by the authors.

\subsubsection{CoNaLa}
The CoNaLa dataset was collected by Yin et al.~\cite{yin2018learning} and consists of 2,879 examples (split into 2,379 training and 500 test examples), crawled from Stack Overflow and then manually curated by human annotators. 
In addition to the main dataset, Yin et al. also provide a large automatically-mined dataset that consists of Stack Overflow ``how to'' questions as training intents and contiguous lines from code blocks in answers as candidate implementations for the intent. 
This dataset has more than a hundred thousand examples. 
Some of the models that we consider use it for training. 
The CoNaLa dataset has the following features:
\begin{itemize}
    \item The CoNaLa dataset has a sound variety of intents that cover many methods used in Python (as compared to, \eg the Card2Code dataset~\cite{card2code}, which is dedicated to the generation of classes with very rigid structure). 
    \item Intents in the CoNaLa dataset are detailed and written in natural language, which distinguishes it from, \eg the Docstrings~\cite{Docstrings} dataset, where the intents are rather short and in many cases a human programmer would have problems with writing the correct code given only the intent. 
    \item There is a relatively rich choice of the publicly available models that were evaluated on this dataset (as compared to the other datasets), enabling us to have more comparisons. 
    \item The best performing models evaluated on the CoNaLa dataset have BLEU scores around 30, allowing to have generated test snippets of both high and low quality. 
    For example, the best model evaluated on the Docstrings dataset has BLEU 12.1, which corresponds to a majority of the snippets being low quality, making it harder for human graders to reliably distinguish between them. 
    \item The CoNaLa snippets are generally very short, with the absolute majority of them being a single line of code. 
    It limits the possible usability of the CodeBLEU and RUBY metrics that take code structure into account.
\end{itemize}

We evaluate five models on the CoNaLa dataset. 
One of the models we consider is the baseline CoNaLa model~\cite{conala}, another is Codex~\cite{chen2021evaluating}, and three others are Transformer-based \texttt{tranX} models. The \texttt{tranx-annot} model was trained on the main CoNaLa dataset; \texttt{best-tranx} was also pretrained on the larger automatically-mined version of CoNaLa before being trained on the main CoNaLa dataset; \texttt{best-tranx-rerank} is the enhanced version of the second model that uses reranking postprocessing (\ie reranking the $n$-best predictions to increase the quality of the output). 
For each of these models, we use the standard setup as provided in the replication package.
Finally, we run Codex~\cite{chen2021evaluating}, specifically, its \texttt{davinci} version, in the Q\&A mode.
Following the authors' recommendations, we do not fine-tune Codex on the CoNaLa training part and rather provide it with three code snippets as examples.
That is, each code snippet is generated via OpenAI Q\&A API for Python code generation, and three intent-snippet pairs are provided as the examples.
It is important to highlight that the exact setup of the models (such as hyperparameter choice or configuration) is not crucial for our study, since we do not try to estimate which model is objectively better, but focus on studying the metrics evaluation of the outputs of code generating models.
Thus, the only non-trivial requirement the outputs should satisfy is that different models should produce snippets of varying quality for the same problem formulation (so that it is possible to create synthetic models of significantly different quality).

\subsubsection{Card2Code Hearthstone}
Card2Code is a pair of datasets derived from the collectible trading card games \textit{Magic the Gathering} and \textit{Hearthstone}; in our research, we focus on the Hearthstone dataset as it is more popular among the researchers. 
The Hearthstone dataset contains 665 pairs of Hearthstone card descriptions and corresponding Python snippets. 
Each snippet is a class implementation that can be used in the Hearthbreaker Hearthstone simulator~\cite{heartbreaker} to describe the card's logic. 
The dataset is split into 533 training pairs, 66 validation pairs, and 66 test pairs. 
The Hearthstone dataset has the following features:
\begin{itemize}
    \item As the intents are the descriptions of Hearthstone cards that should adhere to the Hearthbreaker notation, the generated code has a relatively rigid structure.
    \item The code generation problem is very peculiar: every task requires the model to generate a class. 
    The snippets have very similar outline, and the difference between various snippets is limited: each snippet is a class inherited from one of three parent classes (MinionCard, SpellCard and WeaponCard).
    Almost every snippet has exactly two methods: a constructor and a method with the name depending on the parent class (\textit{use} for SpellCard, \textit{create\_weapon} for WeaponCard).
    Thus, the generality of the conclusions we may infer from the results is limited.
    \item The generated code is relatively long and complex, allowing application of the CodeBLEU and RUBY metrics that take the underlying code structure into account.
\end{itemize}

There are only two publicly available models that are evaluated on the Hearthstone dataset.
One of the models is a syntactic neural model \texttt{NL2code}~\cite{Yin}, and another is a grammar-based structural convolutional neural network \texttt{GCNN}~\cite{GCNN}. 
For \texttt{NL2code}, we use the outputs of the model provided by the authors, and for \texttt{GCNN} we use standard setup as provided in the replication package, but limit training to 30 epochs since the standard setup of 1000 epochs (as written in the replication package) was unfeasible with our computational resources.
The pre-trained Codex model was evidently familiar with the dataset since it provided reference snippets as an output, so we did not consider it.
In particular, without a tight limit on the number of generated tokens, Codex successfully generated several classes from the testing dataset in a single run. 
This suggests that Codex is capable of reproducing entire files that it has seen during training, including the ones from the Hearthstone dataset.

To check the significance of the difference in the metric scores, we use paired bootstrap resampling~\cite{efron1983estimating} for the metric scores of the models evaluated on the test part of the dataset.

\subsection{RQ1: Corpus-level Model Performance}
To address RQ1, we compare the significance of metric score differences on the corpus level.
For the metrics which define corpus-level scores as an aggregate of snippet-level scores, it is possible to use techniques such as Wilcoxon sign-rank test~\cite{wilcoxon1992individual} to compare the models. 
However, there are metrics like BLEU which are corpus-level by design, so that simple averaging of per-snippet scores over the corpus does not give corpus-level metric score (see appendix~\ref{bleu} for more details). 
Thus, the Wilcoxon test is not applicable in this case.
This restricts us to using randomized significance testing for comparing corpus level scores, which is a common practice in the machine translation community~\cite{graham2014randomized}. 
According to Graham et al.~\cite{graham2014randomized}, there is little practical difference between using bootstrap, paired bootstrap and approximate randomization to test significance.
We choose paired bootstrap resampling to test significance.
To test for statistical significance, we take 1000 bootstrap samples.

\subsection{RQ2: Significance of the Automated Metrics' Scores}
To address RQ2, we consider the significance of difference in metric scores for various pairs of the models. 
We expect that the significance of difference in metric scores will vary with the difference in scores (so that for a pair of models with BLEU scores of 20 and 80, it is more likely that one of the models will be better than the other, as compared to the pair of models with BLEU scores of 29.5 and 30).
Thus, we follow Roy et al.~\cite{roy2021reassessing} and split the pairs of models into the bins according to the difference in the scores.
The bin composure ($[0, 1), [1, 2)$ etc.) is slightly different for Hearthstone and CoNaLa dataset. 
It was determined empirically to have similar number of pairs in every bin.
We strive to have a comparable number of pairs in every bin in order to have a significant number of pairs in every bin, so that it is possible to draw statistically robust conclusions. 

We augment our set of original ML models with the synthetic models built according to the approach of Roy et al.~\cite{roy2021reassessing}.
We build the synthetic models' outputs from the outputs of real models.
There are several reasons why we use synthetic models:
\begin{enumerate}
    \item There is a relative scarcity of available models. 
    In the best case of the CoNaLa dataset, we only have five models of various quality, which may not provide enough data to assess the metrics applicability.
    The usage of syntactic models allows us to cover a much more diverse range of metric values without training many new models.
    \item Even if there was a great variety of models so that there would be enough data points for proper metric comparison, it would require immense investment in labeling the data. 
    For example, in this research, we study outputs of 85 models in total (which includes both original and synthetic models) just for the CoNaLa dataset, with each of the outputs consisting of 472 snippets. 
    If all 85 models were independent, it would require people experienced in Python to label more than 40,000 snippets. 
    As we deem it necessary to collect at least three scores for every snippet, such a procedure would be prohibitively hard or expensive.
    \item Improving or worsening the model scores results in a set of synthetic models with the metric and human scores relatively close to each other. 
    This allows us to compare many models with relatively close scores and check the significance of relatively small differences in them.
    This is relevant to the researchers and practitioners, since the improvements over the state-of-the-art models often come in small increments.
\end{enumerate}

\subsubsection{Building Synthetic Models}
We create a synthetic model by starting with the outputs of some of the original models and replacing X\% of its worst-rated snippets with the best-rated snippet for the problem. 
The quality of the snippet is assessed according to the human evaluation scores.
If the picked snippet is already the best-rated snippet, it is skipped.
The reverse procedure is applied for synthetically worsened models.
We continue the replacement procedure until X\% of snippets is changed or there are no more snippets left to change.

Following Roy et al.~\cite{roy2021reassessing}, we consider eight different proportions for the replacements: replacing 1\%, 3\%, 5\%, 10\%, 15\%, 20\%, 25\%, and 30\% of the generated snippets. 
Our replacement proportions are identical to those of Roy et al. with a slight variation: we replace 3\% of the dataset instead of 2\% replaced by Roy et al.
This procedure yields $5 \times 8 \times 2 = 80$ synthetic models for CoNaLa and $2 \times 8 \times 2 = 32$ synthetic models for the Hearthstone dataset. 
Then, we add the original models and deduplicate them by throwing out models with fully identical outputs. This leaves us with 81 models for CoNaLa and 29 models for Hearthstone that we use for our analysis in RQ2.
We consider all pairwise combinations of the models (both synthetic and original) and do the paired difference test for every metric. 

\subsection{RQ3: Agreement Between Metrics and Human Evaluation}
To address RQ3, we assess the degree of agreement between the human assessment and metric scores on the corpus level.
In order to do so, we carry out corpus-level significance tests to check whether the metric and the human prediction agree for every pair of models.
Similarly to the previous research question, we utilize both original and synthetic models we used in RQ2.

\subsubsection{Bins for the Corpus-level Assessment}
There are several options for disagreement between human assessors and a metric for a given pair of models A and B:
\begin{itemize}
    \item When A is better than B according to the metric, but the models are equivalent according to human assessors (Type-I error).
    \item When models A and B are equivalent according to the metric, but one of the models is better according to human assessors (Type-II error).
    \item When model A is better than model B according to the metric, but according to the human assessors, model B is better than model A (Type-I error).
\end{itemize}
We consider all pairwise combinations of the models (both synthetic and original) and do paired difference tests for the human and the metric assessments. 
Using the aggregated human scores as ground truth, we quantify Type-I and Type-II errors of the metric.
As we expect that the probability of a metric to make an error for a pair of models depends on the difference of the models' scores, we divide the data on the metric errors into several bins. 
The \textit{NS} bin corresponds to the cases where the difference in the model scores is insignificant, according to a given metric.
All errors in this bin are Type-I errors.
Other bins correspond to the cases where the difference in the model scores is significant, according to the metric. 
The bin composure for RQ3 is identical to the one we choose for the RQ2.

\subsubsection{Human Evaluation}
To get the human assessment of the considered models, we created a survey, in which we asked programmers to evaluate code snippets. 
The snippets were presented one by one and were randomly chosen out of the combined pool of snippets generated by the models and reference snippets. 
The graders did not know the origin of each snippet.
The graders rated the snippets on the scale from 0 to 4, with the following grade descriptions:
\begin{itemize}
    \item[0:] The snippet is not at all helpful, it is irrelevant to the problem.
    \item[1:] The snippet is slightly helpful, it contains information relevant to the problem, but it is easier to write the solution from scratch.
    \item[2:] The snippet is somewhat helpful, it requires significant changes (compared to the size of the snippet), but is still useful.
    \item[3:] The snippet is helpful but needs to be slightly changed to solve the problem.
    \item[4:] The snippet is very helpful, it solves the problem.
\end{itemize}
The graders did not have to evaluate all snippets in the dataset and could stop at any moment.

\subsubsection{The CoNaLa Dataset}
\begin{figure}[h]
\includegraphics[width=\columnwidth]{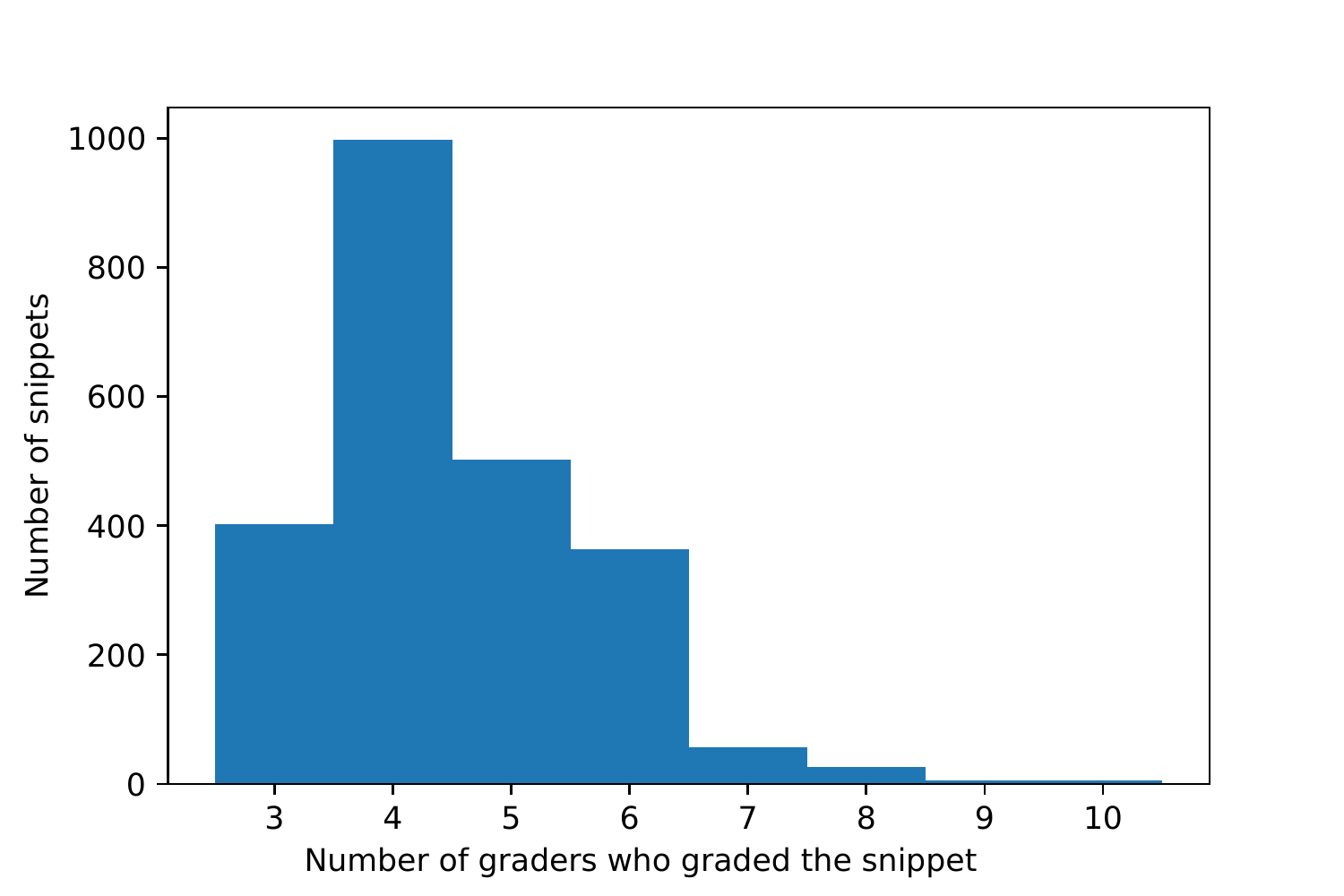} \caption{Distribution of the number of grades per snippet.} \label{fig:grader-distrib}
\end{figure}

For the CoNaLa dataset, there were 2,860 snippets to evaluate: $5\times 472$ snippets generated by the models plus $500$ reference snippets (for some of the intents the dataset contains more than one reference snippet). 
16 participants took part in our survey, and on average, we received 4.49 grades per model-generated snippet. \Cref{fig:grader-distrib} shows the distribution of the number of grades.
Three of the graders have less than two years of experience with Python, six have two to three years of experience, and seven are programming in Python for four or more years. 
We have recruited graders from the ranks of our colleagues and through posts in our scientific Twitter accounts. 
At the moment of grading, all the graders were doing research in the computer science software engineering domain.

\subsubsection{The Hearthstone Dataset}
\begin{figure}[h!]
  \centering
  \theverbbox\qquad
  \includegraphics[scale=.3]{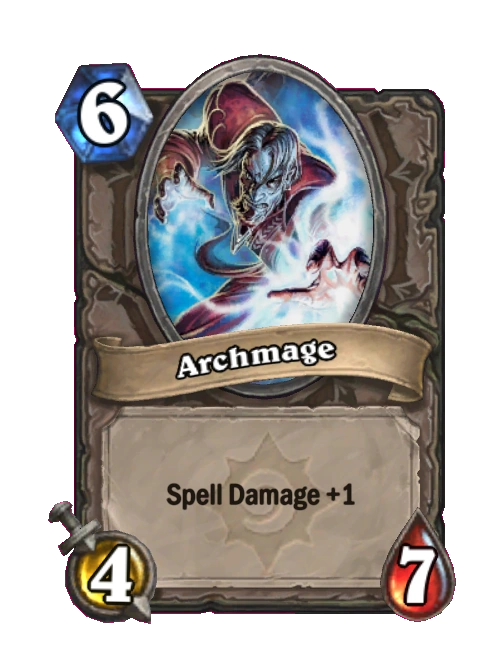}
  \caption{The Archmage card and the corresponding code snippet}\label{fig:archmage}
\end{figure}

Similarly to the CoNaLa dataset, we also ran a survey in which programmers evaluated code snippets. 
The snippets were presented one by one along with the Hearthstone card images, and the graders assessed whether the snippet represents the card correctly. \Cref{fig:archmage} shows an example of a card image along with the corresponding code snippet.

\begin{verbbox}[\mbox{}]
class Archmage(MinionCard ) : 
    def __init__ (self) :
        super().__init__("Archmage", 6, 
        CHARACTER_CLASS.ALL, CARD_RARITY.COMMON)

    def create_minion (self, player) :
        return Minion(4, 7, spell_damage = 1)

\end{verbbox}

There were 198 snippets to evaluate: $2\times 66$ snippets generated by the models plus 66 reference snippets. 
Four participants took part in the survey, every participant has graded all the snippets. 
Two of the participants had three or more years of experience of playing Hearthstone, and two other participants have studied the rules through videos and manuals. 
One of the graders had 1.5 years of experience with Python, two had two years of experience, and one was programming in Python for four years.

\section{Results}\label{sec:results-discussion}

\subsection{RQ1: Corpus-level Model Performance}

\subsubsection{The CoNaLa Dataset}

The test part of the CoNaLa dataset consists of 500 reference snippets, but some of the intents appear more than once, so in total, there are 472 unique intents. 
Different references corresponding to the same intent, were accounted for as parts of the references corpus.
We consider five different models trained on the CoNaLa dataset: baseline CoNaLa (\texttt{baseline}), tranX trained on the main dataset (\texttt{tranx-annot}), the best version of tranX by Yin et al.~\cite{tranx} with pretraining on the non-cleaned version of CoNaLa and without reranking (\texttt{best-tranx}), the best version of tranX with the same pretraining and reranking (\texttt{best-tranx-rerank})~\cite{reranking}, and Codex~\cite{chen2021evaluating}. 
We compute BLEU, ROUGE-L, METEOR, ChrF, CodeBLEU, and RUBY scores for the outputs of these models (getting scores for each of the test snippets). 
\Cref{tab:metrics-conala} shows metric values for all the models on the CoNaLa dataset.

\begin{table*}
    \begin{tabular}{lccccc}\toprule
              & \textbf{baseline} & \textbf{tranx-annot} & \textbf{best-tranx} & \textbf{best-tranx-rerank} & \textbf{Codex} \\ \midrule
         \textbf{BLEU} & $12.37^{+1.59}_{-1.46}$ & $28.58^{+3.18}_{-3.06}$ & $31.48^{+3.01}_{-2.98}$ & $\mathbf{33.14}^{+2.91}_{-2.94}$ & $33.04_{-3.14}^{+3.24}$ \\[0.15cm] 
         \textbf{ROUGE-L} & $36.51^{+1.41}_{-1.46}$ & $49.22^{+1.79}_{-1.69}$ & $51.47_{-1.90}^{+1.87}$ & $52.83^{+1.96}_{-1.84}$ & $\mathbf{56.52}^{+2.25}_{-2.29}$ \\[0.15cm] 
         \textbf{ChrF} & $17.51^{+1.26}_{-1.26}$ & $28.30_{-1.79}^{+1.66}$ & $31.14_{-1.85}^{+1.89}$ & $32.67_{-1.95}^{+2.10}$ & $\mathbf{42.84}_{-2.54}^{+2.68}$ \\[0.15cm] 
         \textbf{METEOR} & $28.43^{+1.54}_{-1.54}$ & $44.03_{-2.03}^{+2.18}$ & $46.55_{-2.30}^{+2.28}$ & $48.32_{-2.38}^{+2.43}$ & $\mathbf{50.66}^{+2.66}_{-2.49}$ \\[0.15cm] 
         \textbf{RUBY} & $43.32^{+1.99}_{-1.84}$ & $43.52^{+1.92}_{-1.93}$ & $44.81^{+2.00}_{-2.00}$ & $46.26^{+2.03}_{-1.97}$ & $\mathbf{57.70}^{+2.24}_{-2.24}$ \\[0.15cm] 
         \textbf{CodeBLEU} & $30.97^{+1.67}_{-1.47}$ & $33.02_{-1.62}^{+1.57}$ & $34.07^{+1.67}_{-1.63}$ & $34.33_{-1.65}^{+1.69}$ & $\mathbf{46.58}_{-2.47}^{+2.64}$ \\[0.15cm] \midrule 
         \textbf{Human} & $8.74_{-1.64}^{+1.80}$ & $26.69^{+3.44}_{-2.91}$ & $35.22_{-3.28}^{+3.23}$ & $40.10_{-3.39}^{+3.70}$ & $\mathbf{59.85}^{+3.50}_{-3.50}$ \\[0.15cm] \bottomrule
    \end{tabular}
    \caption{\label{tab:metrics-conala}Metric results for the CoNaLa dataset.}
\end{table*}

\begin{table*}[t]
\begin{tabular}{lcccccccc}
\toprule
         & \multicolumn{2}{c}{\textbf{Delta: {[}0, 2)}} & \multicolumn{2}{c}{\textbf{Delta: {[}2, 5)}} & \multicolumn{2}{c}{\textbf{Delta: {[}5, 10)}} & \multicolumn{2}{c}{\textbf{Delta: {[}10, 100)}} \\ \cmidrule(lr){2-3} \cmidrule(lr){4-5} \cmidrule(lr){6-7}  \cmidrule(lr){8-9} 
         & \textbf{Significant}    & \textbf{NS}    & \textbf{Significant}    & \textbf{NS}    & \textbf{Significant}     & \textbf{NS}    & \textbf{Significant}      & \textbf{NS}     \\ \midrule
\textbf{BLEU}     & 192            & 398                & 732            & 42                 & 893             & 0                  & 1064             & 0                   \\
\textbf{ROUGE-L}  & 252            & 296                & 736            & 0                  & 1023            & 0                  & 1014             & 0                   \\
\textbf{ChrF}     & 253            & 212                & 633            & 1                  & 922             & 0                  & 1300             & 8                   \\
\textbf{METEOR}   & 195            & 324                & 699            & 7                  & 914             & 0                  & 1182             & 48                  \\
\textbf{RUBY}     & 235            & 437                & 828            & 9                  & 972             & 0                  & 840              & 0                   \\
\textbf{CodeBLEU} & 382            & 474                & 895            & 6                  & 857             & 0                  & 707              & 0                   \\ \hline
\end{tabular}
\caption{\label{tab:significance-conala}Significance for the corpus-level metrics score difference on the CoNaLa dataset.}
\end{table*}

Alongside the automated metrics, we report the aggregated assessor scores (see row \texttt{Human} in \Cref{tab:metrics-conala}). 
We convert all the metrics to the 0--100 scale by multiplying with an appropriate factor: we multiply assessor scores by 25 and multiply automated metric scores by 100, if the metric scores are in the $[0, 1]$ span. 
Together with the scores, we report confidence intervals for each of the metrics.
The confidence intervals were computed with the aid of bootstrap over 1,000 resamplings; $X^{+Y}_{-Z}$ should be read as ``95\% of the resampled models yielded score in the $[X-Z, X+Y]$ range''.

The BLEU metric failed to recognize the difference in quality between Codex and \texttt{best-tranx-rerank}, and between Codex and \texttt{best-tranx}.
The RUBY metric failed to recognize the difference in quality between any of the following three models: baseline, \texttt{tranx-annot}, and \texttt{best-tranx}.
The CodeBLEU metric failed to recognize the difference in quality between any of the two models from the following ones: \texttt{tranx-annot}, \texttt{best-tranx}, and \texttt{best-tranx-rerank} models.
For the five original models evaluated on the CoNaLa dataset the differences in either BLEU, RUBY, or CodeBLEU scores are not always statistically significant.
This is important, since currently code generation models are evaluated with either CodeBLEU or BLEU, and the model scores are often provided without any data on the statistical significance.

\begin{table*}[t]
\footnotesize
\begin{tabular}{lccccccccccc}
\toprule
         & \multicolumn{2}{c}{\textbf{Delta: {[}0, 2)}} & \multicolumn{2}{c}{\textbf{Delta: {[}2, 5) }} & \multicolumn{2}{c}{\textbf{Delta: {[}5, 10)}} & \multicolumn{2}{c}{\textbf{Delta: {[}10, 100)}} & \multicolumn{2}{c}{\textbf{Delta: NS }} & \multirow{2}{*}{\begin{tabular}[c]{@{}c@{}}\textbf{Total} \\ \textbf{mismatch}\end{tabular}} \\ \cmidrule(lr){2-3} \cmidrule(lr){4-5} \cmidrule(lr){6-7} \cmidrule(lr){8-9} \cmidrule(lr){10-11}
         & \textbf{Mismatches}          & \textbf{Pairs}         & \textbf{Mismatches}          & \textbf{Pairs}         & \textbf{Mismatches}          & \textbf{Pairs}          & \textbf{Mismatches}           & \textbf{Pairs}           & \textbf{Mismatches}       & \textbf{Pairs}      &                                                                                    \\ \midrule
\textbf{BLEU}     & 2.7\%               & 187           & 15.1\%              & 747           & 12.0\%              & 890            & 0.6\%                & 1070            & 85.5\%           & 427        & 17.95\%                                                                            \\
\textbf{ROUGE-L}  & 6.7\%               & 254           & 12.0\%              & 740           & 3.7\%               & 1016           & 0                    & 1018            & 72.0\%           & 293        & 10.69\%                                                                            \\
\textbf{ChrF}     & 5.2\%               & 248           & 16.2\%              & 627           & 2.8\%               & 923            & 0                    & 1305            & 64.7\%           & 218        & 8.49\%                                                                             \\
\textbf{METEOR}   & 4.7\%               & 190           & 14.8\%              & 694           & 9.3\%               & 914            & 0                    & 1187            & 81.5\%           & 336        & 14.18\%                                                                            \\
\textbf{RUBY}     & 6.6\%               & 213           & 21.0\%              & 837           & 4.7\%               & 965            & 0                    & 838             & 85.9\%           & 468        & 19.21\%                                                                            \\
\textbf{CodeBLEU} & 6.0\%               & 382           & 9.4\%               & 896           & 5.8\%               & 842            & 0                    & 715             & 80.9\%           & 486        & 16.53\%                                                                            \\ \hline
\end{tabular}
\caption{\label{tab:disagreement-conala}Corpus-level metrics disagreement rate on the CoNaLa dataset.}
\end{table*}

\subsubsection{The Hearthstone Dataset}
For the Hearthstone dataset, we only evaluate two different models available: a syntactic neural model NL2Code~\cite{Yin} and a grammar-based structural convolutional neural network (GCNN)~\cite{GCNN}. 
We compute BLEU, ROUGE-L, METEOR, ChrF, CodeBLEU, and RUBY scores for the outputs of these models, getting scores for each of the test snippets. 
The format in which we report the scores is the same as the format in which we presented CoNaLa scores.
We trained the GCNN model for 30 epochs, as there was no recommended number of epochs in the original paper~\cite{GCNN}, and the default value of 1,000 epochs is unfeasible. 
This may be the reason why the GCNN model we trained performs relatively worse than NL2Code contrary to the results of the original paper~\cite{GCNN}. 

\begin{table}
    \begin{tabular}{lcc}
         \toprule & \textbf{gcnn} & \textbf{nl2code} \\ \midrule
         \textbf{BLEU} & $69.20^{+6.52}_{-6.29}$ & $\mathbf{74.52}^{+6.20}_{-6.03}$ \\[0.15cm] 
         \textbf{ROUGE-L} & $84.71^{+3.53}_{-3.50}$ & $\mathbf{86.54}^{+3.05}_{-3.18}$ \\[0.15cm] 
         \textbf{ChrF} & $\mathbf{80.76}^{+4.21}_{-4.35}$ & $80.60_{-3.78}^{+3.87}$ \\[0.15cm] 
         \textbf{METEOR} & $75.18^{+5.72}_{-5.60}$ & $\mathbf{79.64}_{-4.98}^{+5.35}$ \\[0.15cm] 
         \textbf{RUBY} & $\mathbf{85.82}^{+3.71}_{-3.74}$ & $85.56^{+3.48}_{-3.69}$ \\[0.15cm] 
         \textbf{CodeBLEU} & $71.59^{+6.24}_{-5.84}$ & $\mathbf{72.35}_{-5.57}^{+5.73}$ \\[0.15cm] \midrule
         \textbf{Human} & $65.53_{-6.44}^{+6.44}$ & $\mathbf{68.18}^{+5.68}_{-5.68}$ \\[0.15cm] \bottomrule
    \end{tabular}
    \caption{\label{tab:metrics-hs}Metric results for the Hearthstone dataset.}
\end{table}

\begin{table*}[t]
\begin{tabular}{lcccccccc}
\toprule
         & \multicolumn{2}{c}{\textbf{Delta: {[}0, 1)}} & \multicolumn{2}{c}{\textbf{Delta: {[}1, 2)}} & \multicolumn{2}{c}{\textbf{Delta: {[}2, 4)}} & \multicolumn{2}{c}{\textbf{Delta: {[}4, 100)}} \\ \cmidrule(lr){2-3} \cmidrule(lr){4-5} \cmidrule(lr){6-7} \cmidrule(lr){8-9}
         & \textbf{Significant}    & \textbf{NS}    & \textbf{Significant}    & \textbf{NS}    & \textbf{Significant}    & \textbf{NS}    & \textbf{Significant}     & \textbf{NS}     \\ \midrule
\textbf{BLEU}     & 30             & 91                 & 16             & 56                 & 98             & 16                 & 128             & 0                   \\
\textbf{ROUGE-L}  & 58             & 138                & 67             & 35                 & 137            & 0                  & 0               & 0                   \\
\textbf{ChrF}     & 71             & 134                & 90             & 33                 & 99             & 0                  & 8               & 0                   \\
\textbf{METEOR}   & 21             & 144                & 22             & 33                 & 159            & 8                  & 48              & 0                   \\
\textbf{RUBY}     & 60             & 164                & 76             & 73                 & 62             & 0                  & 0               & 0                   \\
\textbf{CodeBLEU} & 31             & 243                & 22             & 102                & 24             & 13                 & 0               & 0                   \\ \hline
\end{tabular}
\caption{Significance for the corpus-level metrics score difference on the Hearthstone dataset.}
\label{tab:significance-hs}
\end{table*}
\begin{table*}[t]
\footnotesize
\begin{tabular}{lccccccccccc}
\toprule
         & \multicolumn{2}{c}{\textbf{Delta: {[}0, 1)}} & \multicolumn{2}{c}{\textbf{Delta: {[}1, 2}) } & \multicolumn{2}{c}{\textbf{Delta: {[}2, 4)}} & \multicolumn{2}{c}{\textbf{Delta: {[}4, 100)}} & \multicolumn{2}{c}{\textbf{Delta: NS}} & \multirow{2}{*}{\begin{tabular}[c]{@{}c@{}}\textbf{Total}\\ \textbf{mismatch}\end{tabular}} \\ \cmidrule(lr){2-3} \cmidrule(lr){4-5} \cmidrule(lr){6-7} \cmidrule(lr){8-9} \cmidrule(lr){10-11}
         & \textbf{Mismatches}          & \textbf{Pairs}         & \textbf{Mismatches}          & \textbf{Pairs}         & \textbf{Mismatches}          & \textbf{Pairs}         & \textbf{Mismatches}           & \textbf{Pairs}          & \textbf{Mismatches}       & \textbf{Pairs}      &                                                                           \\ \midrule
\textbf{BLEU}    & 3.7\%               & 27            & 0.0\%               & 16            & 24.4\%              & 98            & 27.3\%               & 128            & 81.9\%           & 166        & 45.1\%                                                                    \\
\textbf{ROUGE-L}  & 1.6\%               & 64            & 7.7\%               & 65            & 0.0\%               & 137           &                      & 0              & 50.3\%           & 169        & 20.9\%                                                                    \\
\textbf{ChrF}     & 1.4\%               & 73            & 27.2\%              & 92            & 0.0\%               & 99            & 0.0\%                & 8              & 59.5\%           & 163        & 28.3\%                                                                    \\
\textbf{METEOR}   & 5.9\%               & 17            & 0.0\%               & 22            & 18.9\%              & 159           & 22.9\%               & 48             & 74.6\%           & 189        & 42.1\%                                                                    \\
\textbf{RUBY}     & 2.6\%               & 39            & 4.5\%               & 110           & 3.0\%               & 66            &                      & 0              & 62.7\%           & 220        & 33.6\%                                                                    \\
\textbf{CodeBLEU} & 15.2\%              & 33            & 0.0                 & 22            & 0.0\%               & 24            &                      & 0              & 75.0\%           & 356        & 62.5\%                                                                    \\ \hline
\end{tabular}
\caption{Corpus-level metrics disagreement rate on the HearthStone dataset.}
\label{tab:disagreement-hs}
\end{table*}

According to the ROUGE-L, METEOR, and BLEU metrics, the NL2Code model is better than GCNN with $>95\%$ confidence, see \Cref{tab:metrics-hs}.
\\
\\
\noindent\fbox{%
    \parbox{\linewidth}{
We find that even for the non-synthetic models we consider on the CoNaLa and Hearthstone dataset, the improvement in metric scores may be superficial and statistically insignificant.
This highlights the necessity to test the significance of the improvement in models' quality for the code generation task.
}
}

\subsection{RQ2: Significance of the Automated Metrics' Scores }
\subsubsection{The CoNaLa Dataset}
In \Cref{tab:significance-conala}, we present the data on the significance of differences in model scores.
For every pair of models, we compute the difference in their scores, according to each of the metrics we consider and check whether the difference is significant, according to the paired bootstrap resampling procedure.
Every table cell contains the number of model pairs that correspond to the metric and difference mentioned in the row and column, respectively. For example, there were 192 pairs of models with a difference in BLEU scores in the $[0, 2)$ range, for which this difference was significant.
We split the possible scores into four different bins---$[0, 2)$, $[2, 5)$, $[5, 10)$, $[10, 100)$---and put every pair of models into the corresponding bin.
The results show that with the exception of the BLEU metric, if the difference in metric scores of two models is larger than two points, then it is possible to claim with at least $95\%$ confidence that the difference is significant.
The data on the confidence of the difference significance can be obtained directly from the table: for every bin-metric pair the confidence is given by $S/(S+NS)$. 
Here $S$ is the number of model pairs, for which the difference in scores was significant, and $NS$ is the number of model pairs, for which the difference in scores was not statistically significant.
The results also show that if the difference in scores of two models is less than two points, it is impossible to claim that one of the models is better without carrying out additional statistical tests.
Moreover, if the difference in BLEU scores is less than five points, additional statistical tests are necessary to claim that the difference is significant.

\subsubsection{The Hearthstone Dataset}

\Cref{tab:significance-hs} presents the dependence between the difference in model scores according to the metrics and their ability to determine which model is better with at least 95\% confidence.
The results show that for the Hearthstone dataset, the difference in scores of less than two points according to any metric makes it impossible to claim that one of the models is significantly better without additional statistical tests.
For the adopted by the community BLEU and CodeBLEU metrics---and only for them---it is impossible to claim that one of the models is significantly better if the difference in model scores is less than four points.
Similarly to our results on the CoNaLa dataset, this finding highlights that the small difference in the metric scores should be reported together with the statistical tests that prove the significance of the difference.
\\
\\
\noindent\fbox{%
    \parbox{\linewidth}{
Our findings for the significance of the metric scores improvement extend the observations we made in the previous section. 
We find that for none of the metrics we consider a score improvement of less than two points is sufficient to claim a statistically signficant improvement without additional tests. 
Moreover, the sufficient score improvement to claim a statistically significant improvement for the community-adopted BLEU and CodeBLEU metrics is even higher.
}%
}

\subsection{RQ3: Agreement Between Metrics and Human Evaluation}
\subsubsection{The CoNaLa Dataset}
We also carry out human evaluation of the CoNaLa dataset and compare it with the results of automated metrics.
We computed the ``ground truth'' human grade according to the M-MSR algorithm suggested by Ma et al.~\cite{ma2020adversarial}.

For the non-synthetic models, various metrics show different results in recognizing the significance of difference in the outputs' quality of the models.
The human ground truth we obtained from the collected grades shows that all the differences in the model scores are significant. 
The ranking of the models is as follows: Codex > \texttt{best-tranx-rerank} > \texttt{best-tranx} > \texttt{tranx-annot} > \texttt{baseline}.
The results of the ChrF, ROUGE-L, and METEOR metrics agree with the human judgement (see ~\Cref{tab:metrics-conala}), while BLEU, RUBY, and CodeBLEU disagree with the assessors for at least one pair of models.

We present the comparison of human assessment to the automated metrics on the synthetic models in~\Cref{tab:disagreement-conala}.
Every column contains data on pairs of models with statistically significant difference in the metric scores in the given range.
The \texttt{Delta: NS} column contains all pairs of models, for which the difference in scores was not statistically significant.
Every table cell contains the number of model pairs that correspond to the metric mentioned in the respective row and difference mentioned in the column. For example, there were 187 pairs of models with a difference in BLEU scores in the $[0, 2)$ range, for which this difference was significant.
The ``Mismatches'' column lists the number of model pairs for which the metric assessment disagrees with the human evaluation. For example, out of 187 model pairs with a difference in BLEU scores in the $[0, 2)$ range for which the difference was significant, for 2.7\% the metric assessment did not agree with the human assessment.

For the disagreement rate of the corpus-level metrics with the aggregated human scores, we can see the following:

\begin{itemize}
    \item[1.] The metrics are not reliable in determining that the difference between the models is not significant with an error rate being above 60\% for every metric we consider, see column \texttt{Delta: NS}.
    \item[2.] When the difference in metric scores is less than 5 points, no metric is reliable enough to emulate the human judgement with at least 95\% precision.
    \item[3.] For the $[5, 10)$ bin of the metric scores difference only RUBY, ChrF and ROUGE-L metrics are able to emulate the human judgement with at least 95\% precision, see column \texttt{Delta: {[}5, 10)}.
    \item[4.] It is possible to argue that out of the metrics we consider, BLEU is the worst in emulating human judgement: even though it has the second-highest total mismatch rate, it is the worst-performing metric for the models with a score difference of more than five points, see row \texttt{BLEU}. 
    It is also the only metric that sometimes disagrees with the human judgement for the pair of models that have a score difference of more than 10 points.
    \item[5.] RUBY and CodeBLEU metrics, which were developed for assessing code, do not perform significantly better than the metrics originating from the machine translation domain.
    Moreover, they are among the least reliable in terms of total mismatch rate, see column \texttt{Total mismatch}.
    \item[6.] All metrics have the highest incidence of Type-I errors for the $[2, 5)$ bin, that then decreases with the increase in scores difference, see column \texttt{Delta: {[}2, 5)}. 
    This can be explained by the high mismatch rate in the NS bin, which consists of pairs of models with generally small difference in scores.
    If we do not consider the NS bin separately and aggregate the results according to the difference in pairs of models scores, the highest error rate is for the $[0, 2)$ bin, similarly to the results of Roy et al.~\cite{roy2021reassessing}. 
\end{itemize}

\noindent\fbox{%
    \parbox{\linewidth}{
The general recommendation for the practitioners, based on the results of our study, is that a difference of metric scores of at least five points is necessary to claim with at least 95\% certainty that one model is better than the other on the CoNaLa dataset, if the human judgement is considered to be the golden truth.
ChrF and ROUGE-L are the best-performing metrics for the assessment of code generation models among the metrics we consider.
}}

\subsubsection{The Hearthstone Dataset}
We also conducted the human assessment of the Hearthstone dataset. 
Similarly to CoNaLa dataset, we computed the ``ground truth'' human grade according to the M-MSR algorithm.
For the non-synthetic models, human graders are not able to decide with $>95\%$ confidence that NL2Code is better than GCNN, and the same is true for the CodeBLEU, ChrF, and RUBY metrics.

For the disagreement rate of the corpus-level metrics with the aggregated human scores on the synthetic models, we can see the following:
\begin{itemize}
    \item[1.] The metrics are not reliable in determining that the difference between the models is not significant.
    The relative error rate, however, is slightly better than the one observed for the CoNaLa dataset: ChrF and ROUGE-L exhibit the error rate of less than 60\%, see column \texttt{Delta: NS}.
    \item[2.] The total mismatch rate for the Hearthstone dataset is worse than the one observed for the CoNaLa dataset, see column \texttt{Total mismatch}. 
    The reason for that may be that we only have two models available for the dataset, and their metric scores are relatively close. 
    As all synthetic models were generated from these two, it is not surprising that the synthetic models' scores are also rather close and it is hard for the metrics to discriminate between models.
    \item[3.] None of the metrics is reliable enough to discriminate between the models with a score difference of less than two points with $>95\%$ precision, see column \texttt{Delta: {[}1, 2)}.
    \item[4.] Once again, the BLEU metric performs poorly: its total mismatch rate is among the worst, and, together with METEOR, these are the only two metrics which failed to discriminate well between the models with a score difference of more than two points, see row \texttt{BLEU}.
    \item[5.] RUBY and CodeBLEU, metrics developed for assessing code, do not perform significantly better than the metrics originating from the machine translation domain.
    Moreover, they are among the worst metrics in terms of total mismatch rate.
    \item[6.] There is no clear trend for the Type-I error incidence across all the metrics, unlike it is for the CoNaLa dataset. 
    This can be explained by the bin selection that is different from the one chosen for the CoNaLa dataset.
    Unfortunately, the bin selection similar to the one done for the CoNaLa dataset would be even less informative: for most of the metrics, the bins $[5, 10)$ and $[10, 100)$ would be virtually empty as the two non-synthetic models available for this dataset have relatively close quality.
\end{itemize}

\noindent\fbox{%
    \parbox{\linewidth}{
The general recommendation for practitioners based on the collected results is that a difference of metric scores of at least two points is necessary to claim with at least 95\% certainty that one model is better than the other on the Hearthstone dataset, if the human judgement is considered to be the golden truth.
The ROUGE-L metric is the best-performing metric for the assessment of code generation models on this dataset, with ChrF being the second best.
}}

\section{Study implications}\label{sec:implications}

In this work, we study the applicability of various automated metrics --- BLEU, ROUGE-L, METEOR, ChrF, RUBY, and CodeBLEU --- for evaluation of the code generation models. 

Based on the results, we deduce the following recommendations to the practitioners. 
First, the metric scores should be reported together with the data on the significance of the difference in scores.
Second, the difference in metric scores of less than two points is not enough to claim that one model is better than the other, even if the difference is statistically significant.
Third, despite BLEU and CodeBLEU being the most popular metrics for assessing code generation models, we recommend using ChrF as a standard metric for the code generation tasks. 
We also believe that the community will benefit from a new metric that will be tailored for assessing the code generation task.
In order to support the development of such a metric, we make the collected human assessment scores open-source for both of the studied datasets and encourage other researchers to use them in their work.
Finally, we strongly encourage the practitioners who develop code generation models to publish the outputs of their models, as it is close to impossible to observe small, but significant improvements in code generation without the possibility to carry out statistical tests on both old and new model outputs.

\subsection{Future work}
Using the observations we made above, we see the following directions of future work:

The first is extending this study with other programming languages, datasets, and code generation models.
The obtained data should then be used to assess the applicability of various metrics for the particular dataset and or programming languages.
While such an assessment is costly and long, it would allow comparison of code generation models with greater certainty.
The best course of action would also include full human assessment of the code generation models as it is done for \eg machine translation~\cite{barrault2019findings}.
A particular challenge for labelling dataset would be to collect more than 15 assessments per code snippet. 
According to the findings of Mathur et al.~\cite{mathur2020tangled}, this would allow to assess the quality of various models on the snippet level and track the details of the improvements.

The second is creating new metrics for assessing the quality of code generation models.
Considering the relative success of ML-enhanced metrics for natural language processing tasks~\cite{ma2018results, ma2019results} and for code summarization~\cite{roy2021reassessing}, we surmise that a promising direction for a new metric would be an analog of BERTScore~\cite{bertscore}, that would use embeddings from a large language model to compute the similarity score between the reference and the candidate snippets. 

\section{Threats to Validity}\label{sec:threats-to-validity}
\subsection{External threats}
In this paper, we treat external validity threats as the shortcomings that may affect generalizability of our study to other situations.
First of all, our research is based on two Python datasets: a dataset of Python one-liners and a peculiar Card2Code~\cite{card2code} dataset, for which the models are supposed to generate classes with very specific structure.
It would be interesting to explore other datasets; unfortunately, there is a limited choice of existing datasets, and very few models that can be run on a particular dataset are usually publicly available. 
The most interesting dataset that was left outside the scope of this paper is Docstrings~\cite{Docstrings}. 
Unfortunately, the existing models trained on it perform rather poorly. 
In particular, the best available model has a BLEU score of 12.1~\cite{RUBY}, which means that the expected human grades for its output would be rather poor.

The dataset selection threat is closely related to the model selection threat. 
For every dataset we looked over, except for CoNaLa, there is a relative shortage of available models; in particular, we ran all models that were publicly available for the Hearthstone dataset. 
We contacted the authors of the models that were not open-sourced, but unfortunately got no reply. 
It is possible that different model selection would yield different results.

All the datasets we use have code snippets written in Python. 
While most of the existing public datasets for code generation indeed have code in Python, generation of code in other languages is an important task and the choice of the language might affect the results of a study like ours.

All the external threats to validity are related to the sampling bias issue. 
While we cannot know for sure, if the results of our study hold for other programming languages and other Python datasets, the fact that the popular metrics weakly correlate with human judgement for the studied datasets suggests that it might also be the case for others. Thus, we need an extensive evaluation of code generation metrics to robustly compare models.

\subsection{Internal threats}
In this paper, we consider internal validity threats to be shortcomings that affect the trustworthiness of the causal relationship being tested.
Internal threats to the validity of our study are related to the selection of the human graders.
One of the possible threats is the small average number of grades available per snippet. 
It is possible that due to the limited number of developers who have participated in the evaluation, the human grades we derived are different from the ``true'' human grades for the analyzed snippets.
This issue, unfortunately, is common to many studies which use human assessments. 
The number of human grades we collected per snippet is no less than in the other studies which use human assessment for code~\cite{roy2021reassessing, RUBY}, and our results are in line with the findings of Roy et al.~\cite{roy2021reassessing} studying metrics for code summarization.

A related issue is biased graders. 
A grader may have their own preference in coding style or usage of particular technologies that may affect the grades they assign to the snippets. 
To ameliorate this problem and to follow the standard survey practices~\cite{roy2021reassessing}, we shuffled the presented snippets, and added the correct snippets, so that the graders did not know which snippet is correct or not, in order to smear the possible learning effect across the outputs of different models.

We believe that while all the threats to validity listed above are tangible, we have taken all the necessary measures to mitigate them, and our results are valid and usable for the community.

\section{Conclusion}\label{sec:conclusions}
In this study, we examine the current practice of assessing the quality of code generation models with a single corpus-level score based on automated metrics. In particular, we check whether such evaluation yields statistically significant results and correlates well with the human judgement. 
We consider six metrics---BLEU, ROUGE-L, METEOR, ChrF, CodeBLEU, and RUBY---for code generation models evaluated on two different Python datasets: CoNaLa~\cite{conala} and Hearthstone~\cite{card2code}.

We find that even without taking into account human assessment results, the improvement of a corpus-level metric score by less than 2 points might not be enough to warrant a statistically significant improvement in quality without additional statistical tests.
When we also consider the results of human assessment, we find that for some datasets, even an improvement in score by less than 5 points may not correspond to a statistically significant improvement according to the human judgement. Among the metrics we study, ChrF turns out to be the closest to human assessment. However, it cannot be considered the ``perfect'' metric for code generation and finding such a metric requires further work.

In the future work, we aim to extend our studies to other code generation datasets and models, and carry out a more extensive human assessment that would allow to check the model improvement at the snippet level rather than at the corpus level.

\bibliographystyle{elsarticle-num}
\bibliography{metrics-paper}

\begin{thebibliography}{10}
\expandafter\ifx\csname url\endcsname\relax
  \def\url#1{\texttt{#1}}\fi
\expandafter\ifx\csname urlprefix\endcsname\relax\def\urlprefix{URL }\fi
\expandafter\ifx\csname href\endcsname\relax
  \def\href#1#2{#2} \def\path#1{#1}\fi

\bibitem{balzer1985}
R.~Balzer, A 15 year perspective on automatic programming, IEEE Transactions on
  Software Engineering~(11) (1985) 1257--1268.

\bibitem{chen2021evaluating}
M.~Chen, J.~Tworek, H.~Jun, Q.~Yuan, H.~P. d.~O. Pinto, J.~Kaplan, H.~Edwards,
  Y.~Burda, N.~Joseph, G.~Brockman, et~al., Evaluating large language models
  trained on code, arXiv preprint arXiv:2107.03374 (2021).

\bibitem{Yin}
P.~Yin, G.~Neubig, A syntactic neural model for general-purpose code
  generation, arXiv preprint arXiv:1704.01696 (2017).

\bibitem{GCNN}
Z.~Sun, Q.~Zhu, L.~Mou, Y.~Xiong, G.~Li, L.~Zhang, A grammar-based structural
  cnn decoder for code generation, in: Proceedings of the AAAI Conference on
  Artificial Intelligence, Vol.~33, 2019, pp. 7055--7062.

\bibitem{tranx}
P.~Yin, G.~Neubig, Tranx: A transition-based neural abstract syntax parser for
  semantic parsing and code generation, arXiv preprint arXiv:1810.02720 (2018).

\bibitem{reranking}
P.~Yin, G.~Neubig, Reranking for neural semantic parsing, in: Proceedings of
  the 57th Annual Meeting of the Association for Computational Linguistics,
  2019, pp. 4553--4559.

\bibitem{spider}
T.~Yu, R.~Zhang, K.~Yang, M.~Yasunaga, D.~Wang, Z.~Li, J.~Ma, I.~Li, Q.~Yao,
  S.~Roman, et~al., Spider: A large-scale human-labeled dataset for complex and
  cross-domain semantic parsing and text-to-sql task, arXiv preprint
  arXiv:1809.08887 (2018).

\bibitem{oda}
Y.~Oda, H.~Fudaba, G.~Neubig, H.~Hata, S.~Sakti, T.~Toda, S.~Nakamura, Learning
  to generate pseudo-code from source code using statistical machine
  translation (t), in: 2015 30th IEEE/ACM International Conference on Automated
  Software Engineering (ASE), IEEE, 2015, pp. 574--584.

\bibitem{agashe}
R.~Agashe, S.~Iyer, L.~Zettlemoyer, Juice: A large scale distantly supervised
  dataset for open domain context-based code generation, arXiv preprint
  arXiv:1910.02216 (2019).

\bibitem{card2code}
W.~Ling, E.~Grefenstette, K.~M. Hermann, T.~Ko{\v{c}}isk{\`y}, A.~Senior,
  F.~Wang, P.~Blunsom, Latent predictor networks for code generation, arXiv
  preprint arXiv:1603.06744 (2016).

\bibitem{Docstrings}
A.~V.~M. Barone, R.~Sennrich, A parallel corpus of python functions and
  documentation strings for automated code documentation and code generation,
  arXiv preprint arXiv:1707.02275 (2017).

\bibitem{conala}
P.~Yin, B.~Deng, E.~Chen, B.~Vasilescu, G.~Neubig, Learning to mine aligned
  code and natural language pairs from stack overflow, in: International
  Conference on Mining Software Repositories, MSR, ACM, 2018, pp. 476--486.
\newblock \href {https://doi.org/https://doi.org/10.1145/3196398.3196408}
  {\path{doi:https://doi.org/10.1145/3196398.3196408}}.

\bibitem{lu2021codexglue}
S.~Lu, D.~Guo, S.~Ren, J.~Huang, A.~Svyatkovskiy, A.~Blanco, C.~Clement,
  D.~Drain, D.~Jiang, D.~Tang, et~al., Codexglue: A machine learning benchmark
  dataset for code understanding and generation, arXiv preprint
  arXiv:2102.04664 (2021).

\bibitem{BLEU}
K.~Papineni, S.~Roukos, T.~Ward, W.-J. Zhu, Bleu: a method for automatic
  evaluation of machine translation, in: Proceedings of the 40th annual meeting
  of the Association for Computational Linguistics, 2002, pp. 311--318.

\bibitem{CodeBLEU}
S.~Ren, D.~Guo, S.~Lu, L.~Zhou, S.~Liu, D.~Tang, M.~Zhou, A.~Blanco, S.~Ma,
  Codebleu: a method for automatic evaluation of code synthesis, arXiv preprint
  arXiv:2009.10297 (2020).

\bibitem{RUBY}
N.~Tran, H.~Tran, S.~Nguyen, H.~Nguyen, T.~Nguyen, Does bleu score work for
  code migration?, in: 2019 IEEE/ACM 27th International Conference on Program
  Comprehension (ICPC), IEEE, 2019, pp. 165--176.

\bibitem{roy2021reassessing}
D.~Roy, S.~Fakhoury, V.~Arnaoudova, Reassessing automatic evaluation metrics
  for code summarization tasks, in: Proceedings of the 29th ACM Joint Meeting
  on European Software Engineering Conference and Symposium on the Foundations
  of Software Engineering, 2021, pp. 1105--1116.

\bibitem{efron1983estimating}
B.~Efron, Estimating the error rate of a prediction rule: improvement on
  cross-validation, Journal of the American statistical association 78~(382)
  (1983) 316--331.

\bibitem{ROUGE}
C.-Y. Lin, Rouge: A package for automatic evaluation of summaries, in: Text
  summarization branches out, 2004, pp. 74--81.

\bibitem{METEOR}
S.~Banerjee, A.~Lavie, Meteor: An automatic metric for mt evaluation with
  improved correlation with human judgments, in: Proceedings of the acl
  workshop on intrinsic and extrinsic evaluation measures for machine
  translation and/or summarization, 2005, pp. 65--72.

\bibitem{popovic2015chrf}
M.~Popovi{\'c}, chrf: character n-gram f-score for automatic mt evaluation, in:
  Proceedings of the Tenth Workshop on Statistical Machine Translation, 2015,
  pp. 392--395.

\bibitem{mathur2020tangled}
N.~Mathur, T.~Baldwin, T.~Cohn, Tangled up in bleu: Reevaluating the evaluation
  of automatic machine translation evaluation metrics, arXiv preprint
  arXiv:2006.06264 (2020).

\bibitem{Rabinovich}
M.~Rabinovich, M.~Stern, D.~Klein, Abstract syntax networks for code generation
  and semantic parsing, arXiv preprint arXiv:1704.07535 (2017).

\bibitem{hochreiter}
S.~Hochreiter, J.~Schmidhuber, Long short-term memory, Neural computation 9~(8)
  (1997) 1735--1780.

\bibitem{bengio}
Y.~Bengio, P.~Simard, P.~Frasconi, Learning long-term dependencies with
  gradient descent is difficult, IEEE transactions on neural networks 5~(2)
  (1994) 157--166.

\bibitem{Wei}
B.~Wei, G.~Li, X.~Xia, Z.~Fu, Z.~Jin, Code generation as a dual task of code
  summarization, in: Advances in Neural Information Processing Systems, 2019,
  pp. 6563--6573.

\bibitem{vaswani2017}
A.~Vaswani, N.~Shazeer, N.~Parmar, J.~Uszkoreit, L.~Jones, A.~N. Gomez, L.~u.
  Kaiser, I.~Polosukhin,
  \href{https://proceedings.neurips.cc/paper/2017/file/3f5ee243547dee91fbd053c1c4a845aa-Paper.pdf}{Attention
  is all you need}, in: I.~Guyon, U.~V. Luxburg, S.~Bengio, H.~Wallach,
  R.~Fergus, S.~Vishwanathan, R.~Garnett (Eds.), Advances in Neural Information
  Processing Systems, Vol.~30, Curran Associates, Inc., 2017.
\newline\urlprefix\url{https://proceedings.neurips.cc/paper/2017/file/3f5ee243547dee91fbd053c1c4a845aa-Paper.pdf}

\bibitem{alphacode}
Y.~Li, D.~Choi, J.~Chung, N.~Kushman, J.~Schrittwieser, R.~Leblond, T.~Eccles,
  J.~Keeling, F.~Gimeno, A.~D. Lago, T.~Hubert, P.~Choy, C.~d.~M. d'Autume,
  I.~Babuschkin, X.~Chen, P.-S. Huang, J.~Welbl, S.~Gowal, A.~Cherepanov,
  J.~Molloy, D.~J. Mankowitz, E.~S. Robson, P.~Kohli, N.~de~Freitas,
  K.~Kavukcuoglu, O.~Vinyals,
  \href{https://arxiv.org/abs/2203.07814}{Competition-level code generation
  with alphacode} (2022).
\newblock \href {https://doi.org/10.48550/ARXIV.2203.07814}
  {\path{doi:10.48550/ARXIV.2203.07814}}.
\newline\urlprefix\url{https://arxiv.org/abs/2203.07814}

\bibitem{denkowski2014meteor}
M.~Denkowski, A.~Lavie, Meteor universal: Language specific translation
  evaluation for any target language, in: Proceedings of the ninth workshop on
  statistical machine translation, 2014, pp. 376--380.

\bibitem{leclair2019recommendations}
A.~LeClair, C.~McMillan, Recommendations for datasets for source code
  summarization, arXiv preprint arXiv:1904.02660 (2019).

\bibitem{ma2019results}
Q.~Ma, J.~Wei, O.~Bojar, Y.~Graham, Results of the wmt19 metrics shared task:
  Segment-level and strong mt systems pose big challenges, in: Proceedings of
  the Fourth Conference on Machine Translation (Volume 2: Shared Task Papers,
  Day 1), 2019, pp. 62--90.

\bibitem{ma2018results}
Q.~Ma, O.~Bojar, Y.~Graham, Results of the wmt18 metrics shared task: Both
  characters and embeddings achieve good performance, in: Proceedings of the
  third conference on machine translation: shared task papers, 2018, pp.
  671--688.

\bibitem{liu}
C.-W. Liu, R.~Lowe, I.~V. Serban, M.~Noseworthy, L.~Charlin, J.~Pineau, How not
  to evaluate your dialogue system: An empirical study of unsupervised
  evaluation metrics for dialogue response generation, arXiv preprint
  arXiv:1603.08023 (2016).

\bibitem{reiter2018structured}
E.~Reiter, A structured review of the validity of bleu, Computational
  Linguistics 44~(3) (2018) 393--401.

\bibitem{ma2020adversarial}
Q.~Ma, A.~Olshevsky, Adversarial crowdsourcing through robust rank-one matrix
  completion, arXiv preprint arXiv:2010.12181 (2020).

\bibitem{HCOMP2021/CrowdKit}
D.~Ustalov, N.~Pavlichenko, V.~Losev, I.~Giliazev, E.~Tulin,
  \href{https://www.humancomputation.com/assets/wips_demos/HCOMP_2021_paper_85.pdf}{{A
  General-Purpose Crowdsourcing Computational Quality Control Toolkit for
  Python}}, in: The Ninth AAAI Conference on Human Computation and
  Crowdsourcing: Works-in-Progress and Demonstration Track, HCOMP~2021, 2021.
\newblock \href {http://arxiv.org/abs/2109.08584} {\path{arXiv:2109.08584}}.
\newline\urlprefix\url{https://www.humancomputation.com/assets/wips_demos/HCOMP_2021_paper_85.pdf}

\bibitem{juice}
R.~Agashe, S.~Iyer, L.~Zettlemoyer,
  \href{https://aclanthology.org/D19-1546}{{J}u{IC}e: A large scale distantly
  supervised dataset for open domain context-based code generation}, in:
  Proceedings of the 2019 Conference on Empirical Methods in Natural Language
  Processing and the 9th International Joint Conference on Natural Language
  Processing (EMNLP-IJCNLP), Association for Computational Linguistics, Hong
  Kong, China, 2019, pp. 5436--5446.
\newblock \href {https://doi.org/10.18653/v1/D19-1546}
  {\path{doi:10.18653/v1/D19-1546}}.
\newline\urlprefix\url{https://aclanthology.org/D19-1546}

\bibitem{iyer2018mapping}
S.~Iyer, I.~Konstas, A.~Cheung, L.~Zettlemoyer, Mapping language to code in
  programmatic context, arXiv preprint arXiv:1808.09588 (2018).

\bibitem{yin2018learning}
P.~Yin, B.~Deng, E.~Chen, B.~Vasilescu, G.~Neubig, Learning to mine aligned
  code and natural language pairs from stack overflow, in: 2018 IEEE/ACM 15th
  international conference on mining software repositories (MSR), IEEE, 2018,
  pp. 476--486.

\bibitem{heartbreaker}
earthbreaker -- open source hearthstone simulator,
  \url{https://github.com/danielyule/hearthbreaker}, accessed: 2022-06-06.

\bibitem{wilcoxon1992individual}
F.~Wilcoxon, Individual comparisons by ranking methods, in: Breakthroughs in
  statistics, Springer, 1992, pp. 196--202.

\bibitem{graham2014randomized}
Y.~Graham, N.~Mathur, T.~Baldwin, Randomized significance tests in machine
  translation, in: Proceedings of the Ninth Workshop on Statistical Machine
  Translation, 2014, pp. 266--274.

\bibitem{barrault2019findings}
L.~Barrault, O.~Bojar, M.~R. Costa-Jussa, C.~Federmann, M.~Fishel, Y.~Graham,
  Findings of the 2019 conference on machine translation (wmt19), Association
  for Computational Linguistics (ACL), 2019.

\bibitem{bertscore}
T.~Zhang, V.~Kishore, F.~Wu, K.~Q. Weinberger, Y.~Artzi, Bertscore: Evaluating
  text generation with bert, arXiv preprint arXiv:1904.09675 (2019).

\bibitem{chen2014systematic}
B.~Chen, C.~Cherry, A systematic comparison of smoothing techniques for
  sentence-level bleu, in: Proceedings of the ninth workshop on statistical
  machine translation, 2014, pp. 362--367.

\bibitem{post2018call}
M.~Post, A call for clarity in reporting bleu scores, arXiv preprint
  arXiv:1804.08771 (2018).

\bibitem{rouge-perl}
Rouge 1.5.5 perl script,
  \url{https://github.com/andersjo/pyrouge/tree/master/tools/ROUGE-1.5.5},
  accessed: 2022-05-19.

\bibitem{popovic2016chrf}
M.~Popovi{\'c}, chrf deconstructed: beta parameters and n-gram weights, in:
  Proceedings of the First Conference on Machine Translation: Volume 2, Shared
  Task Papers, 2016, pp. 499--504.

\end{thebibliography}

\appendix
\section{Metrics computation} \label{appendix}
\subsection{BLEU}\label{bleu}
BLEU metric is based on the modified $n$-gram precision measure with a length penalization for the candidate sentences that are shorter than the reference ones. The BLEU score is determined by the following formula:
\begin{equation}
    BLEU = BP \cdot \exp\left(\sum\limits_{n=1}^N w_n \log p_n \right); \quad BP = \min(1, e^{1-r/c}),
\end{equation}
where $BP$ is the brevity penalty with $r$ being the length of the reference and $c$ being the candidate translation length. 
$p_n$ corresponds to the weighted overlap between the bag of n-grams (repeated terms are allowed up to the maximal number of repeats across the references). 
If $S^n_{ref}$ and $S^n_{can}$ are bags of n-grams for the reference and the candidate correspondingly, then 
\begin{equation}
    p_n = \frac{\left|S^n_{ref} \cap S^n_{can}\right|}{\left|S^n_{ref}\right|}
\end{equation}
Finally, $w_n$ are the weights for various n-gram contributions; the standard weights are $w_1 = \ldots = w_4 = \frac14$, $w_{n>4} = 0$.
The original BLEU implementation~\cite{BLEU} is a corpus-level metric, as it accounts for the micro-average precision. 
That is, to compute the precision one has to sum the numerators and the denominators for each hypothesis-reference pair before the division.
It is possible to define sentenceBLEU metric to score individual hypothesis (as is done by, \textit{e.g.}, Roy et al.~\cite{roy2021reassessing}) by considering each hypothesis and references as an independent corpus.
One, however, has to remember that the average of sentenceBLEU over the whole dataset is not necessarily equal to the BLEU evaluated on the dataset.

BLEU values range from 0 to 1, with higher scores corresponding to better n-grams precision. 
However, practitioners often multiply BLEU scores by a factor of 100 in their model quality reports.
The default implementation of the BLEU metric gives zero score to the candidates which have zero overlap in 4-grams with the reference.
This restriction may penalize the candidate sentences of mediocre quality too hard (e.g. for a seven-token reference a candidate that guessed 6 tokens right, but missed the token \#4 will get score zero).
Several smoothing algorithms have been suggested to avoid these situations, a systematic comparison of smoothing techniques for the sentence-level BLEU for the machine translation task can be found in the paper of Chen et al.~\cite{chen2014systematic}

In our study, we use the reference BLEU implementation from the\texttt{sacrebleu} package~\cite{post2018call}.

\subsection{ROUGE-L}
ROUGE-L is a metric from the ROUGE family of metrics first suggested by Lin~\cite{ROUGE}.
It was originally suggested for assessing quality of short text summaries, but then was adopted for other tasks.
The basic notion for the ROUGE-L computation is the longest common subsequence (of hypothesis and reference). 
The common subsequence between two sequences $X = [x_i], Y = [y_j]$ is a sequence $[z_{l}]$ that is a subsequence of both $X, Y$. 
The longest common subsequence is then simply a common subsequence of maximal length.
This allows us to define precision, recall and the ROUGE-L metric for hypothesis $H$ and reference $R$ as
\begin{align*}
    &R_{lcs}(H, R) = \frac{LCS(H,R)}{len(R)} \\
    &P_{lcs}(H, R) = \frac{LCS(H,R)}{len(H)} \\
    &ROUGE_L(H, R) = \frac{(1+\beta^2)P_{lcs}R_{lcs}}{R_{lcs} + \beta^2P_{lcs}}
\end{align*}
$\beta$ is the parameter that determines recall weight, in our evaluation we use $\beta = 1$ (equal weight of precision and recall).
The possible values range from 0 to 1, but similarly to BLEU and other metrics the corpus-level scores are often multiplied by 100 to simplify the perception.
The ROUGE-L is commonly used as a snippet-level metric~\cite{roy2021reassessing}. 
This means that to obtain corpus-level ROUGE-L score one has to average snippet-level scores.
For a simple example, let us consider a reference and two hypothesis:
\begin{itemize}
    \item[$R$]: police killed the gunman  
    \item[$H_1$]: police kill the gunman 
    \item[$H_2$]: the gunman killed police
\end{itemize}
The longest common subsequence between $R, H_1$ is 3 tokens long (first, third and fourth token), and the longest common subsequence between $R, H_2$ is 2 tokens long (either first and second, or third and fourth token).
Thus $ROUGE_L(H_1, R) = 0.75$,  $ROUGE_L(H_2, R) = 0.5$.

We use the implementation of ROUGE-L from the \texttt{rouge-score} package, which yields results identical to the original Perl script~\cite{rouge-perl}.

\subsection{ChrF}
ChrF is an F-measure character-based metric first suggested by Popovic~\cite{popovic2015chrf}. 
It was originally proposed for automatic evaluation of machine translation output. 
As a character-based metric, ChrF doesn't depend on tokenization rules. 
It takes every character into account, except for spaces.
To compute ChrF in its standard definition, one has first to compute character-level precision and recall $chrP_k$, $chrR_k$ for character $k$-grams, where $1 \leq k \leq 6$. 
The total n-gram precision and recall $ChrP$, $ChrR$ is the arithmetical average of $chrP_k$, $chrR_k$ respectively.
Finally, ChrF is computed as 
\begin{equation}
  ChrF\beta  = \frac{(1+\beta^2)ChrP ChrR}{ChrR + \beta^2 ChrP}
\end{equation}
Standard ChrF definition that we use sets $\beta = 2$, as this choice of $\beta$ yields the best results in the machine translation tasks~\cite{popovic2016chrf}.

We use the reference implementation of ChrF from the \texttt{sacrebleu} package.

\subsection{METEOR}
METEOR was created as a metric for machine translation evaluation~\cite{METEOR}.
There are several versions of the metric that have slightly different computation rules. 
In our computations we have used the latest version of the metric -- METEOR 1.5~\cite{denkowski2014meteor}.
Its computation consists of the following steps:
\begin{itemize}
    \item Creating alignment between the hypothesis and the reference strings.
     The alignnment between the hypothesis and the reference strings is a mapping between the unigrams of these strings, such that every unigram in each string maps to zero or one unigrams in the other string.
     The alignment is created in several stages with different rules for the unigram matching in each stage.
     In the first stage, two words are matched if and only if they are identical. 
     In the second stage, they are matched if they are identical after Porter stemming.
     In the third stage, two words are matched if they are synonyms according to the WordNet database.
     Finally, two phrases are matched if they are listed as paraphrases in the corresponding language table.
     The mappings are applied iteratively, and the final alignment is the largest subset of all matches built using the beam search. 
     To determine the final alignment, the following criteria in order of the importance are applied:
     \begin{itemize}
         \item The number of covered words across the both sentences should be maximized.
         \item The number of \textit{chunks} should be minimized. 
         A \textit{chunk} is a contiguous series of matches that has identical ordering in both sentences.
         \item The sum of absolute distances between match start indices in the two sentences should be minimized.
         This is to break ties by preferring to align phrases that occur at similar positions in both sentences.
     \end{itemize}
    \item After the alignment has been built, the words in the hypothesis and reference are split into content and function words according to a special function word list.
    For each of the applied matchers one should count the number of content and function words covered by matches of this type.
    Then one calculates weighted precision and recall $P, R$ using matcher weights and content-function word weight.
    From $P, R$ one then computes the weighted harmonic mean $F_{mean}$.
    Finally, to penalize gaps and differences in the word order, one computes a fragmentation penalty using the total number of matched words and number of chunks.
    The METEOR score is finally computed from $F_{mean}$ and fragmentation penalty.
    
\end{itemize}

We use the implementation of METEOR from the \texttt{sacrerouge} package, which makes use of the original script and provides a Python wrapper for it.

\subsection{RUBY}
The metric is defined as
\begin{equation}
    RUBY(R, C) = \begin{cases}
    GRS(R, C) \qquad \text{if PDGs are applicable}, \\
    TRS(R, C) \qquad \text{if ASTs are applicable}, \\
    STS(R, C) \qquad \text{otherwise} \end{cases}
\end{equation}
Here PDG stands for the program dependence graph and AST stands for the abstract syntax tree, $R$ corresponds to the reference and $C$ corresponds to the candidate. 
$GRS(R, C)$ measures the similarity between two program dependence graphs for $R, C$ as 
\begin{equation}
    GRS(R, C) = 1 - \frac{GED(PDG_R, PDG_C)}{size(PDG_R) + size(PDG_C)},
\end{equation}
where $GED(PDG_R, PDG_C)$ is the edit distance between PDG of the reference code and PDG of the candidate code. 
$size(g)$ is the sum of number of vertices and edges of the graph $g$. $GED(a, b)$ is computed as the minimum number of graph edit operations to transform one graph into another with the allowed graph edit operations on vertexes and edges being insertion, deletion, and substitution. 

In the case the PDG is not available for the candidate snippet, the next fallback option is $TRS(R, C)$, which measures the similarity between the ASTs for the reference and the candidate snippet as
\begin{equation}
    TRS(R, C) = 1 - \frac{TED(AST_R, AST_C)}{size(AST_R) + size(AST_C)},
\end{equation}
where $size(T)$ is the number of nodes in the AST, and $TED(a, b)$ is the edit distance between the ASTs of the reference code $AST_R$ and the candidate code $AST_C$. 
TED is given by the minimum number of the editing operations on the AST nodes (that include addition, deletion, replacement and movement) that make $AST_R$ and $AST_C$ identical. 

Finally, the last fallback option for RUBY, that can always be computed, is the string similarity function $STS(R, C)$ that is defined as
\begin{equation}
    STS(R, C) = 1 - \frac{SED(S_R, S_C)}{max(length(S_R), length(S_C))},
\end{equation}
where $SED(S_R, S_C)$ is the string edit distance between the reference sequence $S_R$ and candidate sequence $S_C$.
It measures the number of token deletion/addition actions the user must make to transform the candidate code into the reference one; $length(t)$ is the length of the sequence $t$. 
Tran et al. motivate this choice of metric by the observation that the more abstract metrics have better correlation with the human judgement.
As Tran et al. do not provide a reference implementation of RUBY, in our study we use our own implementation of the RUBY metric.

\subsection{CodeBLEU}
The CodeBLEU metric as suggested by Ren et al.~\cite{CodeBLEU} is given by 
\begin{align}
    CodeBLEU &= 0.1 \cdot BLEU + 0.1\cdot BLEU_w + \\
    &+0.4 \cdot Match_{ast} + 0.4 \cdot Match_{df},
\end{align}
where: 
\begin{itemize}
    \item BLEU is the usual BLEU metric.
    \item $BLEU_w$ is the BLEU metric computed over unigrams only with keywords given 5 times higher weights. 
    In another words, $BLEU_w$ is a precision for unigrams with BLEU brevity penalty.
    For example, for Python reference \lstinline{for x in lst} and hypothesis \lstinline{for x of} $BLEU_w = e^{-1/3}\frac{6}{12}$.
    \item $Match_{ast}$ is the syntactic AST match. 
    To compute this sub-metric, one first has to build the AST for both reference and hypothesis, and extract all sub-trees from both ASTs.
    To track the syntactic structure, authors disregard the values in the leave nodes. 
    $Match_{ast}$ is then given by $Match_{ast} = Count_{clip}(T_{hyp}) / Count(T_{ref})$, where $Count(T_{ref})$ is the total number of sub-trees in reference AST and $Count_{clip}(T_{cand})$ is the number of sub-trees in hypothesis AST that are matched by sub-trees in the reference.
    \item $Match_{df}$ is the semantic data-flow match that considers the semantic similarity between the hypothesis and the reference by comparing the data-flow graphs of the reference and the hypothesis.
    The sub-metric is computed in several steps as follows:
    \begin{itemize}
        \item[1.] Build the data-flow graph for the reference and the hypothesis.
        To do that, one first has to get the variable sequence $V = \{v_0, v_1, \ldots, v_m\}$ from the AST.
        Each variable then becomes a node of the data-flow graph, and directed edges $\epsilon = \langle v_i, v_j \rangle$ signify that the value of $j$-th variable comes from the $i$-th variable.
        The graph $G = (V, E)$ is the data-flow graph.
        \item[2.] Normalize data-flow items.
        To do that, one has to collect all the variables in the data-flow items and rename them $var_i$, where $i$ is the order of the variable appearance in all data-flow items.
        \item[3.] Calculate the semantic data-flow match score as $Match_{df} = Count_{clip}(DF_{hyp}) / Count(DF_{ref})$, where $Count(DF_{ref})$ is the total number of the reference data-flows and \\$Count_{clip}(DF_{cand})$ is the number of the matched candidate data-flows.
    \end{itemize}
\end{itemize}

Ren et al. compared CodeBLEU with BLEU and accuracy.
As CodeBLEU was not compared to other automatic metrics apart from BLEU and accuracy, we need to carry out further assessment.
In our study, we use our own implementation of the CodeBLEU metric.

\end{document}